\documentclass[aps,pre,twocolumn,groupedaddress]{revtex4-1}

\usepackage[english]{babel}
\usepackage[utf8]{inputenc} % Required for inputting international characters
\usepackage[T1]{fontenc} % Output font encoding for international characters

\usepackage{graphicx}
\usepackage{amssymb}
\usepackage{amsmath}
\usepackage{physics} % To write easy physics equations, insert [italicdiff] to make the differential d's italic
\usepackage{siunitx}
\usepackage{bm}%
\usepackage{natbib}
\usepackage[colorlinks=true,linkcolor=blue]{hyperref}%
\usepackage{subeqnarray}
\usepackage{graphics}
\usepackage{pstricks}
\usepackage{amsbsy}
\usepackage{xcolor}

\renewcommand{\vec}{\mathbf}

\newcommand{\la}{\langle}
\newcommand{\ra}{\rangle}

\newcommand{\bRa}{{\bf R}_{\rm a}}

\newcommand{\bRb}{{\bf R}_{\rm b}}

\newcommand{\be}{\begin{equation}}
\newcommand{\ee}{\end{equation}}
\newcommand{\fig}[1]{Fig.~\ref{#1}}
\newcommand{\Fig}[1]{Figure~\ref{#1}}

\newcommand{\eq}[1]{Eq.~(\ref{#1})}
\newcommand{\Eq}[1]{Equation~(\ref{#1})}

\begin{document}

\title{{\black Generalized} hydrodynamics of the Lennard-Jones liquid in view of hidden scale invariance}

\author{Solvej Knudsen$^{1,2}$}
\email{solvejk@ruc.dk}
\author{B. D. Todd$^2$} 
\author{Jeppe C. Dyre$^1$}
\author{J. S. Hansen$^1$}
\affiliation{$^1$
``Glass and Time'', IMFUFA, Department of Science and Environment, Roskilde University, Postbox 260, DK-4000 Roskilde, Denmark 
}
\affiliation{$^2$
Department of Mathematics, School of Science, Computing and Engineering Technologies, Swinburne University of Technology, Hawtorn, Victoria 3122, Australia
}

\begin{abstract}
In recent years lines along which structure and dynamics are invariant to a good approximation, so-called isomorphs, have been identified in the thermodynamic phase diagrams of several model liquids and solids. This paper reports computer simulations of the transverse and longitudinal collective dynamics at different length scales along an isomorph of the Lennard-Jones system. Our findings are compared to corresponding results along an isotherm and an isochore. Confirming the theoretical prediction, the reduced-unit dynamics of the transverse momentum density is invariant to a good approximation along the isomorph on all time and length scales. Likewise, the wave-vector dependent shear-stress autocorrelation function is found to be isomorph invariant (with minor variations at very short times). A similar invariance is not seen along the isotherm or the isochore. Using a spatially non-local hydrodynamic model for the transverse momentum-density time-autocorrelation function, the macroscopic shear viscosity and its wave dependence are determined, demonstrating that the shear viscosity is isomorph invariant on all length scales studied. This analysis implies the existence of a novel length scale that is isomorph invariant in reduced units, i.e., which characterizes each isomorph. The transverse sound-wave velocity, the Maxwell relaxation time, and the rigidity shear modulus are also isomorph invariant. In contrast to the isomorph invariance of all aspects of the transverse dynamics, the reduced-unit dynamics of the mass density is not invariant on length scales longer than the inter-particle distance. By fitting to a generalized hydrodynamic model, we extract values for the wave-vector-dependent thermal diffusion coefficient, sound attenuation coefficient, and adiabatic sound velocity. The isomorph variation of these quantities in reduced units on long length scales can be eliminated by scaling with the density-scaling exponent, a fundamental quantity in the isomorph theory framework, an empirical observation that remains to be explained theoretically.
\end{abstract}

\maketitle

\newpage

\section{Introduction}\label{sec:intro}
Hydrodynamics describes the macroscopic flow of {\black gases} and liquids in terms of continuous time- and space-dependent fields, notably those of mass, momentum, and energy \cite{Landau_Lifshitz_1959,batchelor_introduction_1967}. In the classical treatment the equations expressing conservation of these quantities are supplemented by linear constitutive relations \cite{Groot_Mazur_1984}. In this work we focus on the transverse momentum and mass-density autocorrelation functions, as these contain all relevant information of classical hydrodynamics \cite{Boon_Yip_1991}. The shear-stress autocorrelation function is shown as well, as it gives a different representation of the same information as the transverse momentum density. {\black Alley and Alder carried out a similar analysis for hard spheres back in the 1980s \cite{AlleyAlder1983}.} 

During the last decade it has become clear that many model liquids, including the Lennard-Jones (LJ) system, have the approximate symmetry ``hidden scale invariance''. A consequence of this is that the thermodynamic phase diagram becomes effectively one-dimensional because it has curves, termed isomorphs, along which structure and dynamics are invariant to a good approximation in reduced units \cite{Bailey_Pedersen_Gnan_Schrder_Dyre_2008a,Bailey_Pedersen_Gnan_Schrder_Dyre_2008b,Schrder_Bailey_Pedersen_Gnan_Dyre_2009,Gnan_Schrder_Pedersen_Bailey_Dyre_2009,Schrder_Gnan_Pedersen_Bailey_Dyre_2011}. Hidden scale invariance expresses that the ordering of configurations according to their potential energy is maintained if these are scaled uniformly to a different density. If the position vector of all $N$ particles is denoted by $\vb R\equiv (\vb r_1,...,\vb r_N)$ and $U(\vb R)$ is the potential-energy function, hidden scale invariance is the following logical implication \cite{Schrder_Dyre_2014}
\be\label{HSI}
U(\bRa)<U(\bRb)\,\Rightarrow\,U(\lambda\bRa)<U(\lambda\bRb)\,.
\ee
Here $\lambda$ quantifies the uniform scaling. Hidden scale invariance applies rigorously only for the unrealistic case of an Euler homogeneous potential-energy function (plus a constant) like, e.g., that of a purely repulsive inverse power-law pair potential. \Eq{HSI} applies to a good approximation, however, for the LJ system and its generalizations to mixtures and to exponents other than 6-12 \cite{Bailey_Pedersen_Gnan_Schrder_Dyre_2008a,Pedersen_Schrder_Dyre_2018,Friisberg_Costigliola_Dyre_2017}, to the Yukawa pair-potential system \cite{Veldhorst_Schrder_Dyre_2015}, the exponential repulsive EXP system \cite{Bacher_Schrder_Dyre_2018a}, etc. Interestingly, some molecular models like the Wahnstrom OTP model or the flexible LJ chain model also obey hidden scale invariance and have isomorphs \cite{veldhorst_scaling_2014}. A system with hidden scale invariance is termed ``R-simple'' to distinguish it from the classical definition of a ``simple'' pair-potential system \cite{hansen_mcdonald_2013} (certain pair-potential systems like the Dzugutov system or the Gaussian core model are not R-simple, while some molecular models as mentioned are). 

An isomorph is by definition a curve of constant excess entropy, i.e., an isomorph is a configurational adiabat \cite{Gnan_Schrder_Pedersen_Bailey_Dyre_2009}. While all systems have configurational adiabats, however, only R-simple systems have isomorphs. By now isomorph theory has been applied to many different systems in simulations, but there are also experimental confirmations of isomorph-theory predictions \cite{roed_communication_2013,Gundermann_Pedersen_Hecksher_Bailey_Jakobsen_Christensen_Olsen_Schrder_Fragiadakis_Casalini,Hansen_Sanz_Adrjanowicz_Frick_Niss_2018,XIAO2015190}. Recent reviews of the isomorph theory are given in Refs. \onlinecite{dyre_simple_2016,dyre_perspective_2018}.

The isomorph-theoretical framework has been applied to liquids, glasses, and crystals, but almost all validations of isomorph invariance of the dynamics have focused on single-particle properties like the time-dependent mean-square displacement and the incoherent intermediate scattering function. The collective properties previously considered briefly are the heat conductivity and the shear and bulk viscosities {\black \cite{Costigliola_Schrder_Dyre_2016, heyes_transport_2019}}. Of these the first two were found to be isomorph invariant to a good approximation for the LJ system, whereas the bulk viscosity was not. This paper presents the first systematic investigation of hydrodynamics from the isomorph-theory perspective. We give results for the generalized hydrodynamics, i.e., on several length scales, and investigate how the isomorph-invariance depends on the length scale. We have chosen to study the Lennard-Jones (LJ) system because it is the standard model of liquid-state theory. 

Isomorph invariance is never exact for realistic systems. This means that one cannot expect the hydrodynamic characteristics of the LJ system to be absolutely invariant along the system's isomorphs. In order to be able to judge the degree of invariance, we therefore compare the variation of generalized hydrodynamics properties along an isomorph with those along an isotherm with the same density variation, as well as along an isochore (curve of constant density) with the same temperature variation. Most results are presented in two unit systems, the standard units of molecular dynamics (MD) and the so-called reduced units that depend on the thermodynamic state point (see below), the unit system in which isomorph invariance of structure and dynamics is generally predicted.

\section{Theoretical methods}\label{sec:theoryandsim}
Details of the MD simulations are given below, followed by an introduction to the isomorph theory's reduced units. Hereafter we review the definitions of the hydrodynamic autocorrelation functions (ACFs) studied numerically. Finally, some necessary background of the isomorph theory is given.

\subsection{Simulation details}
MD simulations are carried out using RUMD \cite{bailey_rumd_2017}. We study the standard 12-6 LJ pair-potential which depends on a characteristic energy $\varepsilon$ and length $\sigma$. If $r$ is the distance between two particles, the LJ pair potential $v(r)$ is defined as
\begin{equation}
    v(r) = 4\varepsilon \left[ \left(\frac{\sigma}{r}\right)^{12} - \left(\frac{\sigma}{r}\right)^6 \right]\,.
\end{equation} 
The simulations are carried out in the $NVT$ ensemble with $N$ denoting the number of particles, $V$ the volume, and $T$ the temperature. The thermostat used is Nosé-Hoover. Each simulation involves 6800 particles in a cubic box with side length $L$ and periodic boundary conditions. The potential is truncated and shifted at $r=2.5$ (MD units). Each simulation runs for $10^7$ time steps with each step equal to $0.005$ MD time units. {\black The equilibration are negligibly short compared to the total simulation time and are therefore not excluded from the analysis.}

Data for the time-autocorrelation functions are averaged over $5000$ independent initial configurations and calculated as a Fourier series with wave vectors given by $k = 2\pi p / L$, where $p$ is the wave number and $L$ is the box length. It is costly to simulate many wave vectors, so the simulations were split into two categories: 1) Frequent sampling (every second time step) and a total of 10 wave vectors to ensure a high resolution even for small times. These simulation data are used to investigate the transverse autocorrelation functions directly. 2) Less frequent sampling (every fifth time step) and a total of 50 wave vectors, providing data with a lower resolution but a larger spectrum. These data are used for calculating the viscosity kernel as well as {\black for investigating the} slower longitudinal dynamics.

\subsection{Two unit systems: Transitioning to dimensionless quantities}
For computer simulations it is customary to cast all quantities in so-called MD units. Following Allen and Tildesley \cite{allen_computer_2017} we now list the relevant quantities, where no star denotes the quantity in question while a star denotes the same quantity made dimensionless by reference to MD units. Let $l$ be length, $m$ mass, $t$ time, and $E$ energy. Then the MD units are based on $\sigma$ and $\varepsilon$ from the LJ potential, leading to the following MD dimensionless quantities
\begin{align}
l^* &= l/\sigma \quad \qquad \qquad, \,\,\,m^*=1\\ 
t^* &= \left(\frac{\varepsilon}{m\sigma^2}\right)^\frac{1}{2} t \qquad, \,\,\,E^* = E/\varepsilon\,.
\end{align}

In the reduced units of isomorph theory, the length unit is derived from the particle number density, $n\equiv N/V$, and the energy unit is the thermal energy $k_BT$. This leads \cite{Gnan_Schrder_Pedersen_Bailey_Dyre_2009} to the following dimensionless quantities
\begin{eqnarray}
\tilde{l} = n^{1/3} l\qquad \qquad \qquad &,&\,\,\,  \tilde{m}=1\\
\tilde{t} = n^{1/3} \left({k_B T}/{m}\right)^{1/2} t\,\,\,&,& \,\,\,\tilde{E} = E/k_B T\,.
\end{eqnarray}
Here and henceforth a tilde denotes a dimensionless reduced quantity in the above isomorph-theory sense. 

A state point's number density and temperature is below reported in MD units because these two quantities are both unity in reduced units. For simplicity, the rest of the paper omits the stars when a quantity is given in MD units.

\subsection{Generalized hydrodynamic relaxation functions}
{The collective hydrodynamics are studied} through space and time correlations of the transverse momentum density and shear stress {(transverse dynamics)}, as well as the mass density (longitudinal dynamics). These quantities are defined in terms of the microscopic variables of a computer simulation by the equations given below.

The mass density $\rho(\vb r,t)$ is defined by the atomic masses $m_j$ by \cite{hansen_mcdonald_2013} 
\begin{equation}\label{eq:rho}
    \rho(\vb r,t) = \sum_j m_j \delta(\vb r - \vb r_j(t))
\end{equation}
with $\vb r_j$ being the position of the \textit{jth} particle. From the mass balance equation, the momentum density $\vb j(\vb r,t) = \rho(\vb r, t)\vb u(\vec r,t)$ can be defined as (see \cite{hansenNEWbook} for further explanation)
\begin{align}\label{eq:rho_u_r}
\rho(\vb r, t)\vb u(\vec r,t) = \sum_{j} m_j \vb v_j(t) \delta(\vec r-\vec r_j(t))\,,
\end{align}
where $\vb u(\vb r,t)$ is the mass average velocity and $\vb v_j$ the single particle velocity. The next step in developing hydrodynamics is to imagine the two above expressions averaged in space over a volume that is small enough to allow for studying spatial variations but large enough to contain many particles. From this perspective, one first writes the local mass density and mass average velocity in terms of average and fluctuating parts,
\begin{equation}
    \rho = \rho_{0} + \delta \rho \qq{and} \vb u = \delta \vb u\,,
\end{equation}
since the average streaming velocity is zero. Keeping only terms to first order in the fluctuations, the momentum density reads in Fourier space \cite{Hansen_Dyre_Daivis_Todd_Bruus_2015}
\begin{equation}
    \rho_0\, \delta \vb u(\vb k,t) = \sum_j m_j \vb v_j(t) e^{- i \vb k \vdot \vb r_j(t)}\,.
    \label{eq:rho:u:k}
\end{equation}
Without loss of generality this can be simplified by choosing $\vb k$ to be parallel to one of the coordinate axes, e.g., $\vb k = (0, 0, k)$. Choosing the velocity perpendicular to this, $\delta u_x(k,t) $, the transverse momentum density is given by \cite{Hansen_Dyre_Daivis_Todd_Bruus_2015}
\begin{equation}
    \rho_0 \delta u_x(k, t) = \sum_j m_j v_{j,x}(t) e^{- i k r_{j,z}(t)}.
    \label{eq:rho:ux:k}
\end{equation}
\Eq{eq:rho:ux:k} is used in the simulations to calculate the wave-vector-dependent transverse momentum-density time-autocorrelation function (TMACF) $C_{uu}(k,t)$ defined by \cite{Hansen_Dyre_Daivis_Todd_Bruus_2015}
\begin{align}
	C_{uu} ( k,t) &= \frac{\rho_0^2}{V}\, \la \delta u_x(k, t) \delta u_x(- k,0) \ra\,.
	\label{eq:tacf:k}
\end{align}
Here the angle brackets denote an ensemble average over independent initial conditions, which in practice is replaced by a sample average.

\begin{figure}
\centering
\includegraphics[width=0.4\textwidth]{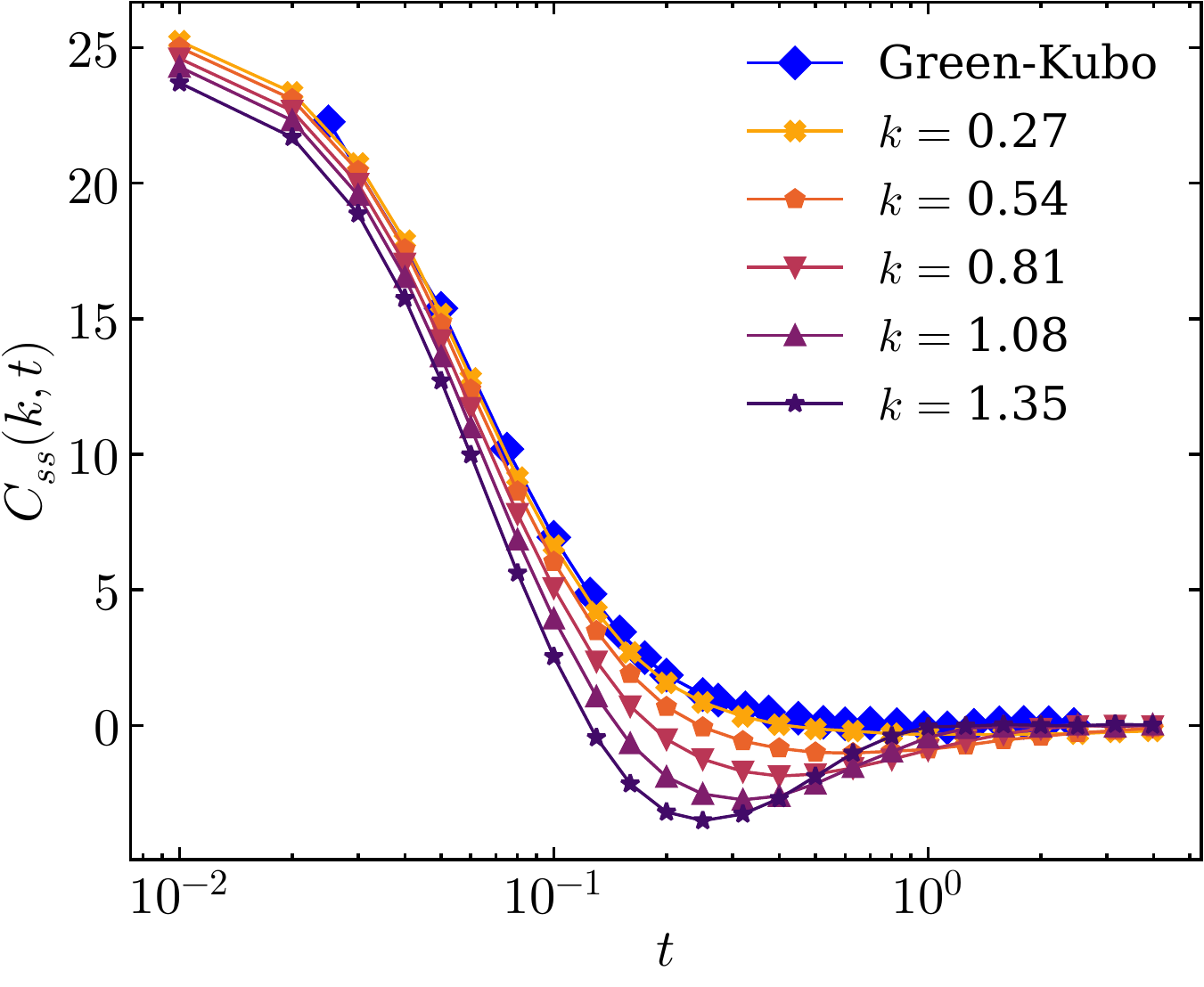}
\caption{A comparison between the Green-Kubo autocorrelation for the shear stress \eq{eq:sacf} (blue points) and the wave vector dependent autocorrelation derived in the text \eq{eq:sacf:k} for a range of wave-vectors (other colors). We see that the data for the wave-vector-dependent stress approach the macroscopic-method data for decreasing $k$. The data shown are for the state point $(n,T)=(0.8,1.1)$.}
\label{fig:sacf:GK}
\end{figure}

Considering next the wave-vector-dependent stress time-autocorrelation function (SACF), the below derivation is straightforward though not standard (see Todd and Daivis \cite{Todd_Daivis_book} for the standard approach). The starting point is the momentum balance equation in Fourier space. If $P_{zx}$ is the $zx$ component of the pressure tensor (the negative stress tensor), the balance equation reads to lowest order in the density fluctuations for $\vb k = (0, 0, k)$
\begin{equation}
\label{eq:momentbalance}
    \rho_0 \pdv{t}\var u_x(k, t) = - i k P_{zx}(k,t).
\end{equation}
Substituting the expression for the transverse momentum density \eq{eq:rho:ux:k} into the balance equation \eq{eq:momentbalance} one obtains (for non-zero $k$)
\begin{equation}
    P_{zx}(k,t) = \sum_{j} \qty[ i \frac{F_{j,x}(t)}{k} + m_j v_{j,z}(t) v_{j,x}(t) ] e^{-i k r_{j,z}(t)}\,.
    \label{eq:Pzx:k}
\end{equation}
Here $F_{j,x}$ is the $x$ component of the total force on particle $j$. This expression is used to form the wave-vector-dependent transverse SACF $C_{ss}(k,t)$ defined by %%CITE THIS FROM WHERE?
\begin{align}
	C_{ss} ( k,t) &= \frac{1}{V} \la P_{zx}( k, t) P_{zx}(- k,0) \ra. 
	\label{eq:sacf:k}
\end{align}
To illustrate this and validate our simulations, we have compared the simulation data at the state point $(n,T)=(0.8,1.1)$ with data for the time-autocorrelation function of the standard expression for the macroscopic, i.e., spatially averaged, pressure tensor's off-diagonal component 
\begin{align}
	C_{ss} (t) &= \frac{1}{V} \la P_{zx}(t) P_{zx}(0) \ra,
	\label{eq:sacf}
\end{align}
with the Irving-Kirkwood expression for the pressure tensor \cite{irving_kirkwood_1950}
\begin{equation}
    P_{zx}(t) = \frac{1}{V}\left(\sum_{j}  r_{j,z}(t)F_{j,x}(t) +  m_j v_{j,z}(t) v_{j,x}(t)\right) \,.
    \label{eq:Pzx:k=0}
\end{equation}
As seen in \fig{fig:sacf:GK}, the data for the wave-vector-dependent SACF nicely approaches the standard macroscopic method data for decreasing values of $k$.

Below, data are given for both the wave-vector-dependent SACF and TMACF as functions of time. Although these two quantities by \eq{eq:momentbalance} are not independent, we report them both because they focus on different aspects of the dynamics.

Turning now to the longitudinal dynamics, the Fourier transform of \eq{eq:rho} gives the wave-vector-dependent mass density,
\begin{align} 
\label{eq:rho_u_k}
\rho(\vec k,t) &= \sum_{j} m_j e^{- i\vec k \cdot \vec r_j(t)}.
\end{align}
This expression is used in the simulations to compute the mass-density time-autocorrelation function (DACF), defined by \cite{Hansen_Dyre_Daivis_Todd_Bruus_2015}
\begin{align}
C_{\rho \rho}(\vec k, t) &= \frac{1}{N}\la \rho(\vec k, t) \rho(- \vec k, 0)\ra\,.
\end{align}
If this quantity is divided by $m^2$ where $m$ is the LJ particle mass, one arrives at the number density time-autocorrelation function.

The DACF is related to the coherent intermediate scattering function $F(k,t)$ \cite{hansen_mcdonald_2013} by
\begin{align}
F(k,t) &=  \frac{1}{n} C_{\rho \rho}(k, t)/m^2\,.
\label{eq:Fkt:DACF}
\end{align}
Since the static structure factor $S(k)$ is defined as
\begin{equation}
S(k) = F(k,t=0),
\label{eq:Sk:Fkt}
\end{equation}
it is possible to obtain $S(k)$ from the DACF data. More generally, one gets the dynamic structure factor $S(k, \omega)$ from the coherent intermediate scattering function by performing a Fourier-Laplace transformation \cite{hansen_mcdonald_2013}
\begin{align}
S(k, \omega) &= \int_0^{\infty} F(k, t) e^{-i\omega t} \dd{t}.
\end{align}

\subsection{Isomorph theory}
The purpose of this paper is to investigate to what degree the hydrodynamics of the LJ system are invariant along an isomorph in the thermodynamic phase diagram. Isomorphs are present whenever the system in question has a high correlation between its potential energy $U$ and virial $W$ constant-volume thermal-equilibrium fluctuations \cite{pedersen_strong_2008, Bailey_Pedersen_Gnan_Schrder_Dyre_2008b}. A measure of the correlation is given by the standard Pearson correlation coefficient, $R$, defined by %Jesper thinks we repeat ourselves
\begin{equation}
R = \frac{\la \Delta U \Delta W \ra}{\sqrt{\la (\Delta U)^2 \ra} \sqrt{\la(\Delta W)^2 \ra} }.
\end{equation}
The correlation is considered to be high whenever $R>0.9$ \cite{Bailey_Pedersen_Gnan_Schrder_Dyre_2008a}. 

\begin{figure}
	\centering
	\includegraphics[width=0.45\textwidth]{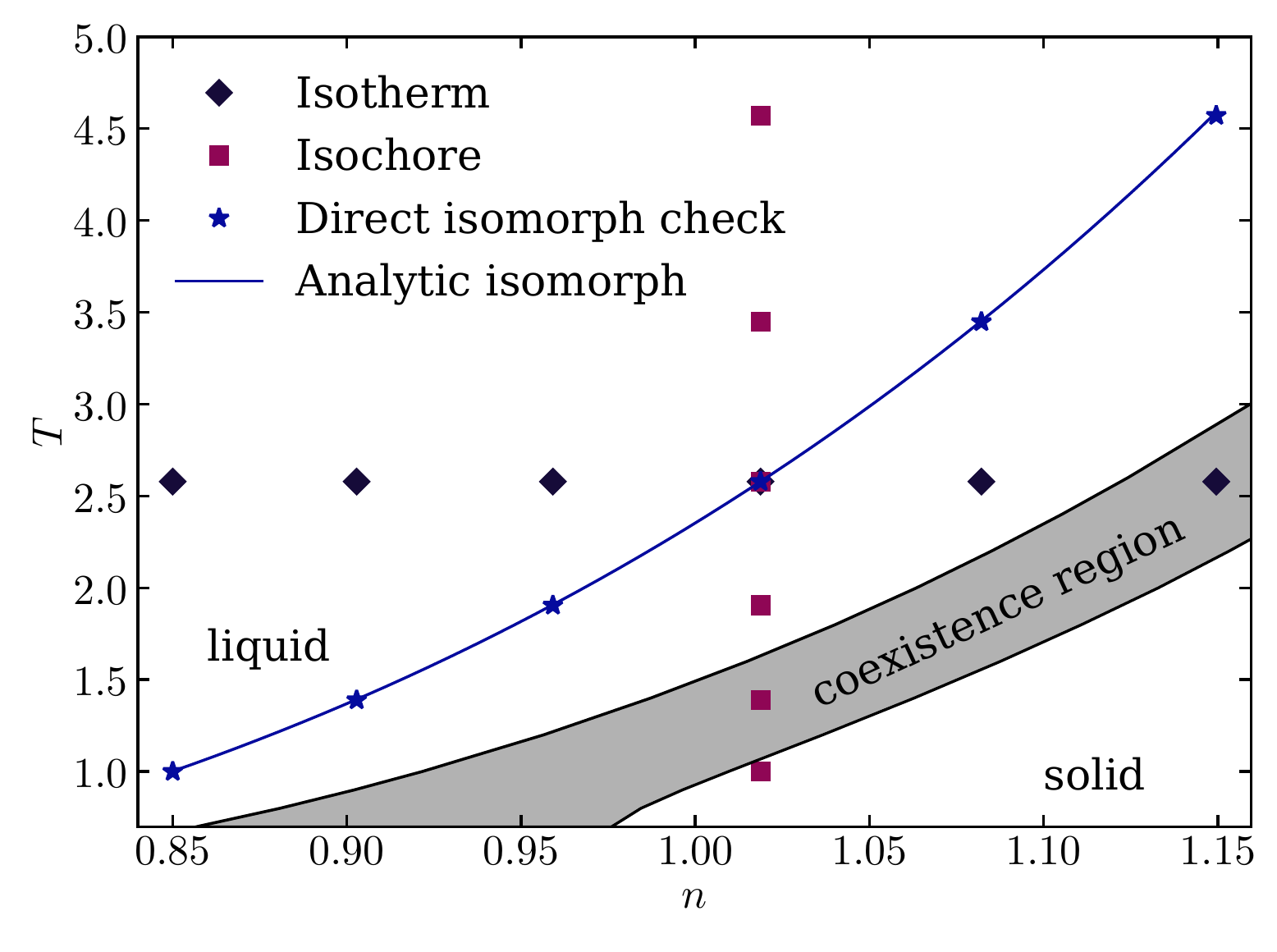}
	\caption{Phase diagram of the Lennard-Jones system with an isomorph, an isotherm, and an isochore through the reference state point $(n_{0}, T_0)=(1.02,2.58)$. For identifying the isomorph, both the direct-isomorph-check method and the analytical formula (\eq{eq:analytical:isomorph}) were used, giving consistent results. Data for the melting and freezing lines were generated from the analytical formula derived in Ref. \cite{pedersen_thermodynamics_2016}. The points in {\black the figure mark} the state points simulated.}
	\label{fig:LJphasediagram}
\end{figure}

The isomorph theory is approximate for all but inverse-power-law systems (for which $R=1$), implying that exact isomorph invariance is not expected. In order to put into perspective the degree of isomorph invariance, we compare below the hydrodynamics along the isomorph with those along an isotherm and isochore. In \fig{fig:LJphasediagram} the LJ thermodynamic phase diagram is shown with the studied isomorph (blue points and curve), isotherm (black points), and isochore (red points). These curves intersect at the ``reference'' state point $(n, T)=(1.02, 2.58)$. Note that the isotherm and isochore span, respectively, the same density and temperature variations as the isomorph. For each state point, data for TMACF, SACF, and DACF have been obtained through MD simulations. The solid black lines are the melting and freezing lines \cite{pedersen_thermodynamics_2016} and the grey area is the coexistence region. The two state points on the isochore with the lowest temperature and the isotherm state point with the highest density are situated below the freezing line. No crystallization was observed at these supercooled state points, however, and we believe it to be safe to include them in the analysis.

The isomorph was traced out using the so-called direct-isomorph-check method starting from the state point $(n,T)=(0.85,1.00)$. In this numerical method a predicted linear relationship between potential energies of scaled and unscaled configurations is utilized in a step-by-step fashion to find temperatures of isomorphic state points. The method works as follows \cite{Gnan_Schrder_Pedersen_Bailey_Dyre_2009}. Let again $U$ be the potential energy, $\vb R$ the position vector of all particles, and $T$ the temperature. Let moreover subscripts $1$ and $2$ refer to two configurations that scale uniformly into one another, corresponding to the densities $n_1$ and $n_2$, respectively. According to the isomorph theory, if the state points $(n_1,T_1)$ and $(n_2,T_2)$ are on the same isomorph, the potential energy of configuration $2$ is given  \cite{Gnan_Schrder_Pedersen_Bailey_Dyre_2009} by
\begin{equation}\label{DIC}
U(\vb R_2) \approx \frac{T_2}{T_1} U(\vb R_1) + D_{12}\,.
\end{equation}
Here $D_{12}$ is a constant offset. In our simulations, configuration $2$ is obtained by increasing the density of configuration $1$ by $1\%$. By computing the potential energy of both configurations and inserting the temperature $T_1$ of the initial configuration, \eq{DIC} determines $T_2$ from a scatter plot of the potential energies of scaled versus unscaled configurations \cite{Gnan_Schrder_Pedersen_Bailey_Dyre_2009}. The virial potential-energy correlation coefficients are above 0.97, see Tables \ref{tab:R:gamma:isomorph}, \ref{tab:R:gamma:isotherm}, and \ref{tab:R:gamma:isochore} for all values of $R$.

An analytical method for tracing out an isomorph of the LJ system exists, which we checked against the direct-isomorph-check method. In \fig{fig:LJphasediagram} results from the analytical method are shown as the solid blue curve while the direct-isomorph-check results are the blue points. The analytical method was derived in Ref. \cite{ingebrigtsen_communication_2012,bohling_scaling_2012} and requires only one simulation at a single state point. In terms of the density-scaling exponent $\gamma$ defined  \cite{Gnan_Schrder_Pedersen_Bailey_Dyre_2009} as (where $\Delta$ denotes a deviation from the thermal average value)
\begin{equation}
\gamma = \frac{\la \Delta U \Delta W \ra }{ \la (\Delta U)^2 \ra},
\label{eq:gamma}
\end{equation}
the analytical formula for the isomorph of state points $(n,T(n))$ that includes the reference state point $(n_0,T_0)$, is given by (where $\gamma_0$ is the density-scaling exponent at the reference state point)
\begin{equation}
\frac{T(n)}{T_0} = \left(\frac{\gamma_0}{2}-1\right)\left(\frac{n}{n_{0}}\right)^4 - \left(\frac{\gamma_0}{2}-2\right)\left(\frac{n}{n_{0}}\right)^2.
\label{eq:analytical:isomorph}
\end{equation}
The first term of \eq{eq:analytical:isomorph} derives from the $r^{-12}$ repulsive term of the LJ potential and the second term derives from its $r^{-6}$ attractive term. At the reference state point $(n_{0}, T_0)=(1.02,2.58)$ we find $\gamma_0=4.91$. Values of $R$ and $\gamma$ at all state points simulated are given in Tables \ref{tab:R:gamma:isomorph}, \ref{tab:R:gamma:isotherm}, and \ref{tab:R:gamma:isochore}.
\begin{table}
\centering
    \caption{Virial potential-energy correlation coefficient $R$ and density-scaling exponent $\gamma$ for the simulated state points along the isomorph.}
\begin{tabular}{ p{1cm} p{1cm} p{1cm} p{1cm} }
$n$ & $T$ & $\gamma$ & R \\
 \hline
 \hline
0.85 & 1.00 & 5.62 & 0.970 \\
0.90 & 1.39 & 5.34 & 0.985 \\
0.96 & 1.90 & 5.11 & 0.992 \\
1.02 & 2.58 & 4.91 & 0.995 \\
1.08 & 3.45 & 4.77 & 0.997 \\
1.15 & 4.57 & 4.65 & 0.998 \\

\end{tabular}
\label{tab:R:gamma:isomorph}
\end{table}
\begin{table}
\centering
    \caption{$R$ and $\gamma$ along the isotherm.}
\begin{tabular}{ p{1cm} p{1cm} p{1cm} p{1cm} }
 $n$ & $T$ & $\gamma$ & R \\
 \hline
 \hline
0.85 & 2.58 & 5.18 & 0.987 \\
0.90 & 2.58 & 5.10 & 0.990 \\
0.96 & 2.58 & 5.00 & 0.993 \\
1.02 & 2.58 & 4.91 & 0.995 \\
1.08 & 2.58 & 4.83 & 0.996 \\
1.15 & 2.58 & 4.74 & 0.997 \\
\end{tabular}
\label{tab:R:gamma:isotherm}
\end{table}
\begin{table}
\centering
    \caption{$R$ and $\gamma$ along the isochore.}
\begin{tabular}{ p{1cm} p{1cm} p{1cm} p{1cm} }
 $n$ & $T$ & $\gamma$ & R \\
 \hline
 \hline
1.02 & 1.00 & 5.16 & 0.992\\
1.02 & 1.39 & 5.07 & 0.993\\
1.02 & 1.90 & 4.99 & 0.994\\
1.02 & 2.58 & 4.91 & 0.995\\
1.02 & 3.45 & 4.84 & 0.995\\
1.02 & 4.57 & 4.78 & 0.996\\
\end{tabular}
\label{tab:R:gamma:isochore}
\end{table}

\section{Results for the transverse hydrodynamics}

This section presents the simulation results for the wave-vector-dependent transverse (shear) stress time-autocorrelation function (SACF) and the wave-vector-dependent transverse momentum time-autocorrelation function (TMACF). 

\begin{figure*}
\centering
\includegraphics[width=0.85\textwidth]{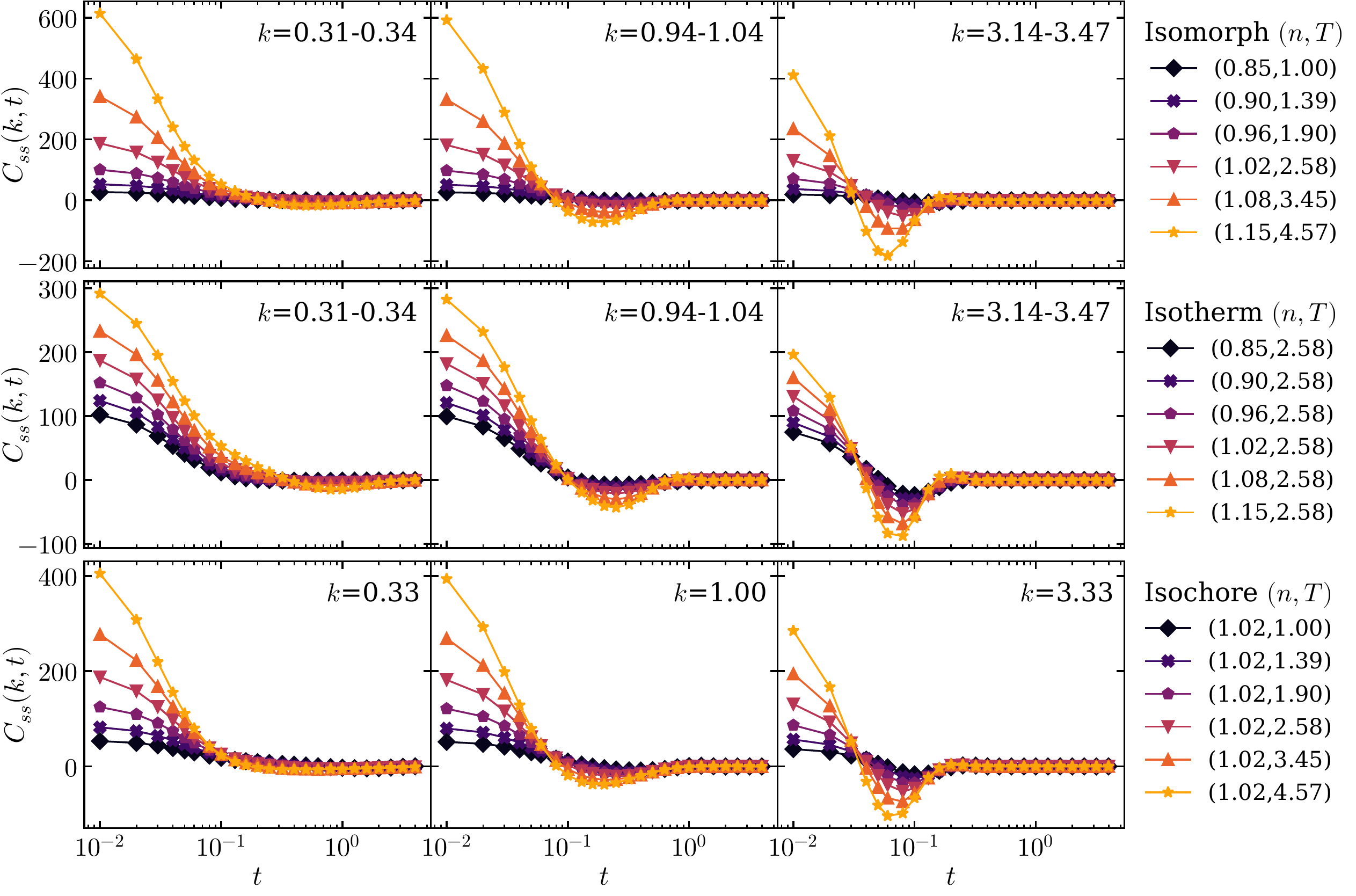}
\caption{Wave-vector-dependent transverse (shear) stress time-autocorrelation function in MD units for three wave vectors ($k\sim 1/3$, $k\sim 1$, $k\sim 3$; constant in reduced units) along the isomorph (top row), isotherm (middle row), and isochore (bottom row). }
\label{fig:sacf:MD}
%\end{figure*}

 \vspace*{\floatsep}%

%\begin{figure*}
%\centering
\includegraphics[width=0.85\textwidth]{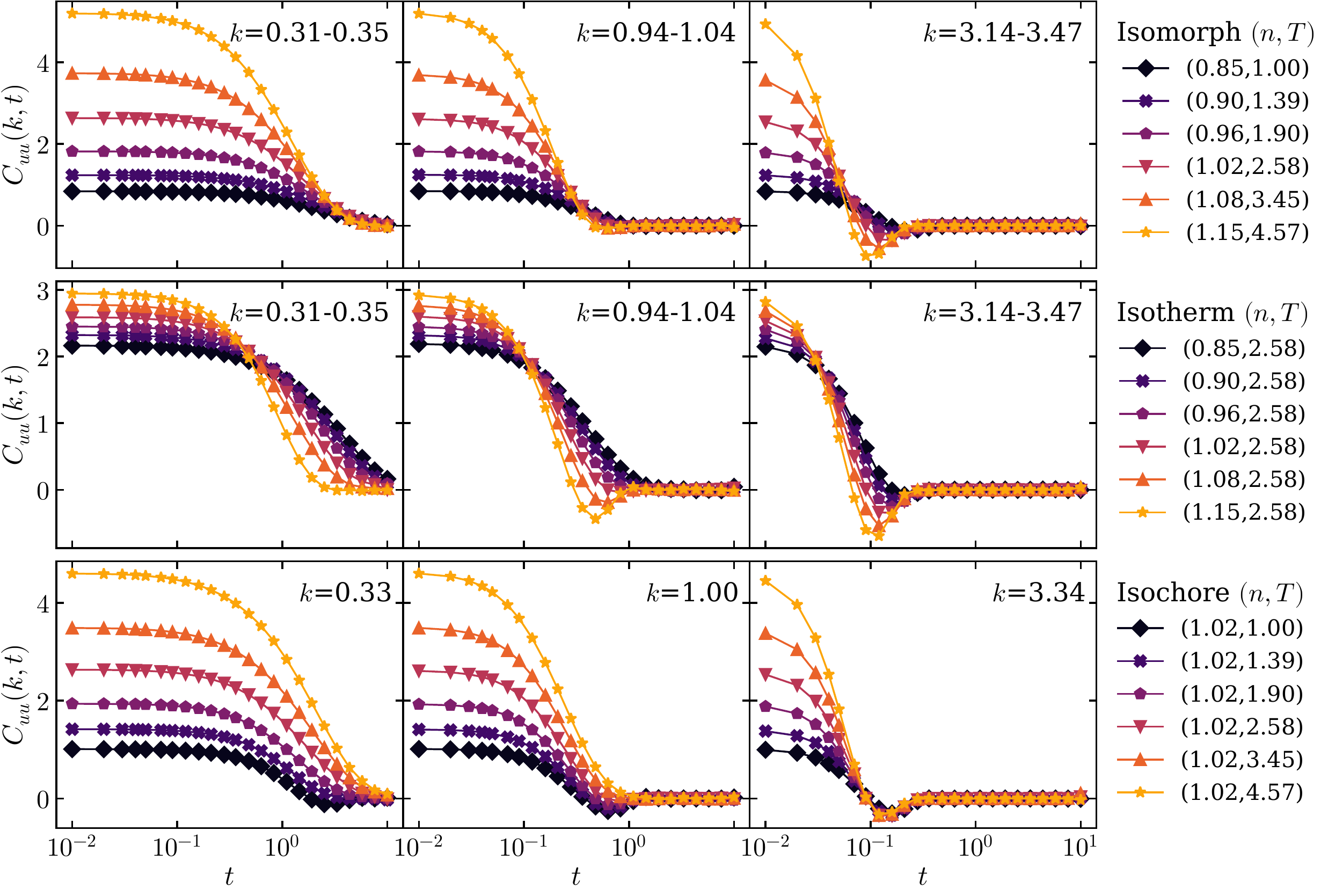}
\caption{Wave-vector-dependent transverse momentum density time-autocorrelation function in MD units for three wave vectors ($k\sim 1/3$, $k\sim 1$, $k\sim 3$) along the isomorph (top row), isotherm (middle row), and isochore (bottom row).}
\label{fig:tacf:MD}
\end{figure*}

\subsection{Transverse stress and momentum-density time-autocorrelation functions}

The wave-vector-dependent transverse SACF is shown in MD units in \fig{fig:sacf:MD}, which is organized with the top row giving results along the isomorph, the middle row along the isotherm, and the bottom row along the isochore. The wave vectors were selected to be identical in reduced units, which is why in MD units they vary across the compared state points (except along the isochore). 

The columns correspond to increasing wave vectors, showing data for $k\sim 1/3$, $k\sim 1$, and $k\sim 3$, respectively. The data all relax to zero at long times, as expected. In the short-time region, the differences between the data for the first and last state points are significant in all three cases, decreasing with increasing wave vector. The shape of the SACF changes significantly as $k$ increases and anti-correlations begin to appear. These are signatures of viscoelastic properties of the liquid, which disappear in the classical hydrodynamic limit ($k \rightarrow 0$). Note that anti-correlations are present already for $k\sim 1/3$ for the supercooled state point on the isotherm (yellow stars).

The wave-vector-dependent TMACF is shown in MD units in \fig{fig:tacf:MD}, where the arrangement of the figure is the same as in \fig{fig:sacf:MD}. Not all $k\sim 1/3$ TMACF data  relax to zero at long times within the simulated time window; we address the consequence of this later. Apart from the tail at long times for the higher wave vectors, the data vary over the full time range. In contrast to the SACF, the variance does not decrease with increasing wave vector. Again, anti-correlations appear for large wave vectors (and for the black squares in the isochore for $k=0.33$ referring to a supercooled state point).

\begin{figure*}
\centering
\includegraphics[width=0.85\textwidth]{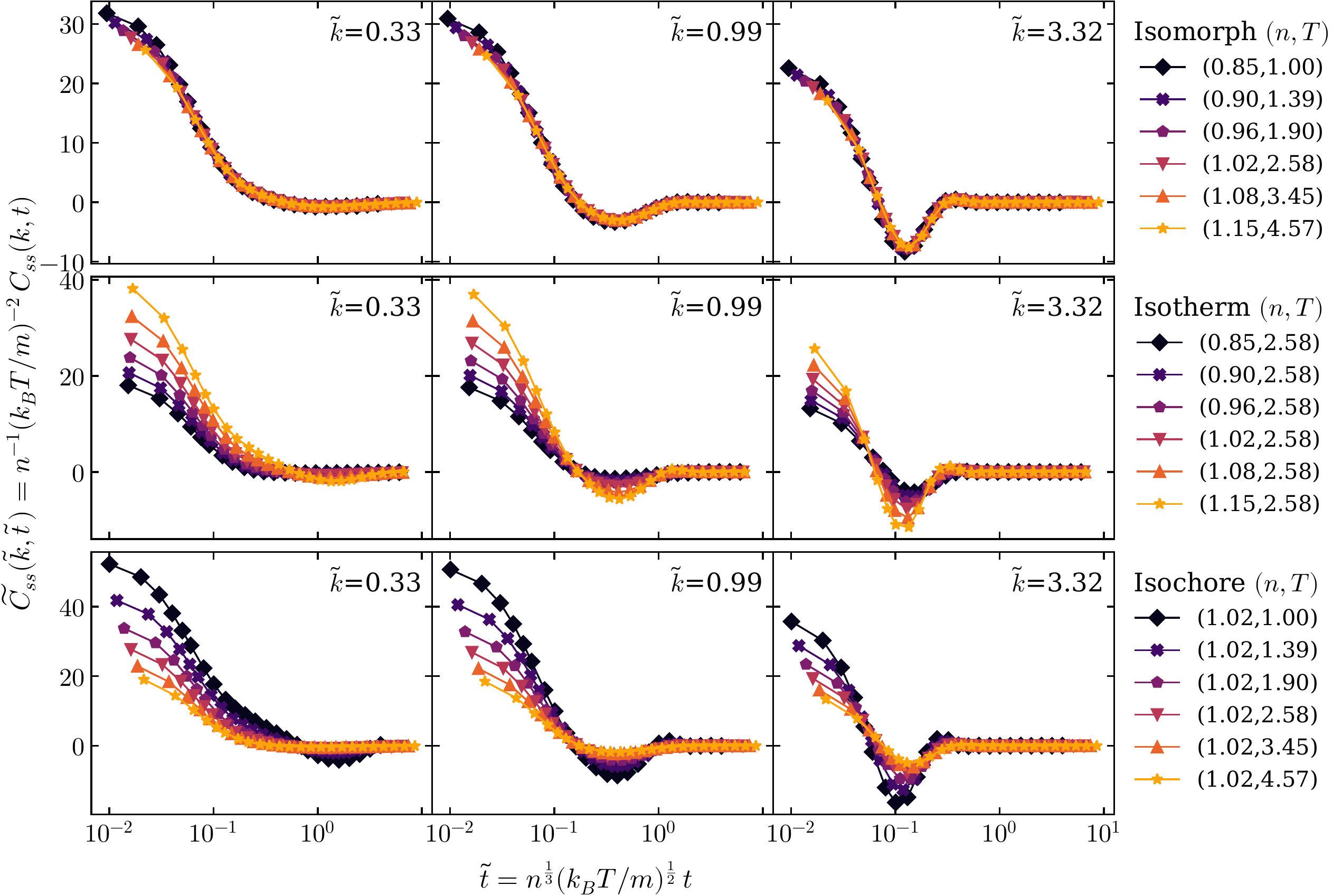}
\caption{Replotting in reduced units the data of Fig. 3 for the Wave-vector-dependent transverse stress time-autocorrelation function for the three wave vectors $\tilde{k}=0.33$, $\tilde{k}=0.99$, and $\tilde{k}=3.32$ along the isomorph (top row), isotherm (middle row), and isochore (bottom row). }
\label{fig:sacf:MR}
%\end{figure*}

 \vspace*{\floatsep}%

%\begin{figure*}
%\centering
\includegraphics[width=0.85\textwidth]{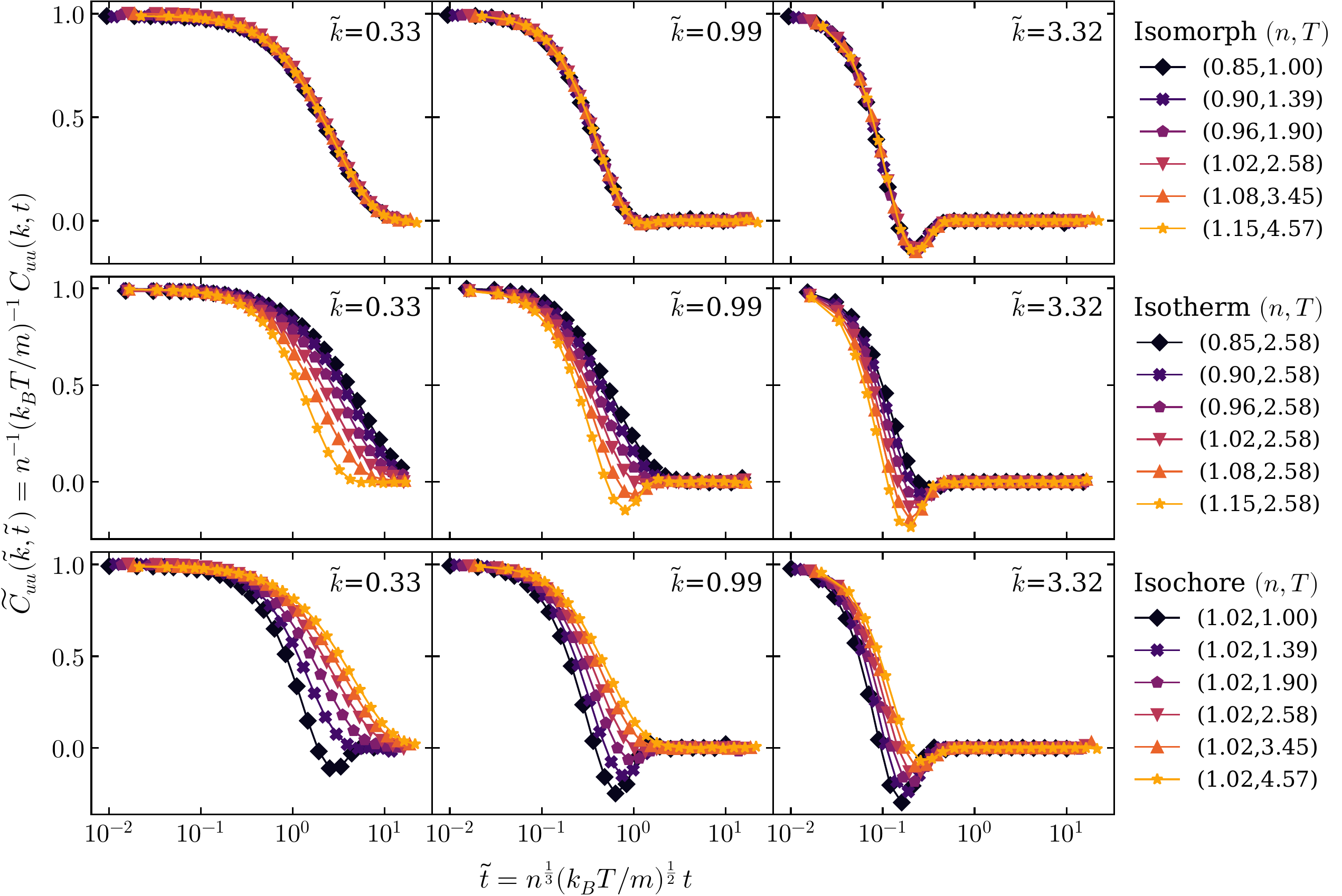}
\caption{Replotting in reduced units the data of Fig. 4 for the Wave-vector-dependent transverse momentum density time-autocorrelation function for the three wave vectors $\tilde{k}=0.33$, $\tilde{k}=0.99$, $\tilde{k}=3.32$ along the isomorph (top row), isotherm (middle row), and isochore (bottom row). }
\label{fig:tacf:MR}
\end{figure*}

Considering now the reduced-unit wave-vector-dependent transverse SACF in \fig{fig:sacf:MR}, the top (isomorph) row shows data collapsing onto single curves. These data are the only ones that exhibit invariance; in fact the isomorph data are invariant on all length scales studied. Even the anti-correlations collapse, demonstrating that the viscoelastic part of the response is also isomorph invariant. Although the deviations from collapse along both the isotherm and isochore decrease compared to when plotted in MD units, variations are still much larger than those of the isomorph. In the short-time region the isomorph does show some variation; this illustrates the fact that the isomorph theory is not exact. 

For the reduced-unit wave-vector-dependent TMACF we also see an excellent collapse along the isomorph at all wave-vectors and at all times (\fig{fig:tacf:MR}). Note that all data sets of \fig{fig:tacf:MR} approach unity in the limit $t \rightarrow 0$; this is a consequence of the equipartition theorem that implies $C_{uu}(k,t\!=\!0) = \rho_0 k_B T$, which in reduced units (putting $k_B=1$) becomes unity. Except for this common short-time limit, the isotherm and isochore data vary considerably in contrast to those of the isomorph. This shows that the observed isomorph collapse is not a consequence of the use of reduced units.

\section{Analysis of the transverse hydrodynamics}

\subsection{Classical hydrodynamics}
\begin{figure}
\centering
\includegraphics[width=0.4\textwidth]{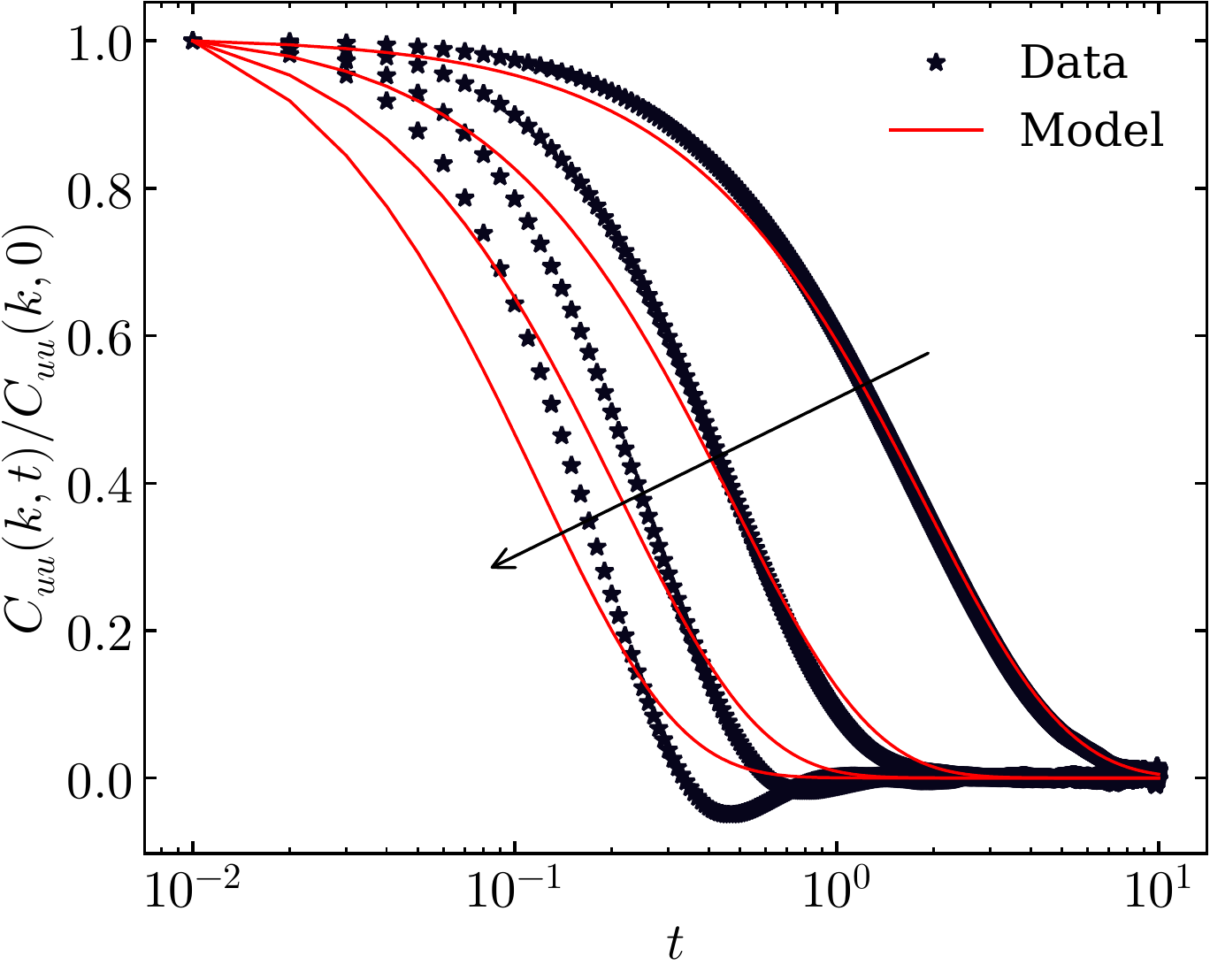}
\caption{The points are TMACF data for $k$ ranging from 0.33 to 1.33 where the arrow indicates the direction of increasing $k$ . The data are taken at the reference state point $(n_0,T_0)=(1.02,2.58)$. The lines are the prediction of classical hydrodynamics (\eq{eq:tacf:exp}) with $\eta_0=4.84\pm0.06$ (found by fitting \eq{eq:fit:fct} to the data in \fig{fig:viscositykernel:1}).}
\label{fig:tacf:exp}
\end{figure}

We first compare the predictions of classical hydrodynamics to the data. For our problem Newton's law of viscosity is
\begin{equation}
    P_{zx}=-\eta_0\, \frac{\partial}{\partial z}  \delta u_x\, .
\end{equation}
In Fourier space this reads $P_{zx}(k,t)=-ik\eta_0\delta u_x(k,t)$, which when substituted into Eq. (\ref{eq:momentbalance}) leads to
\begin{equation}
    \rho_0\,\frac{\partial}{\partial t}\delta u_x(k,t) = -\eta_0\,k^2\,\delta u_x(k,t) \, . 
\end{equation}
Multiplying by $\delta u_x(-k,0)$ and ensemble averaging one gets
\begin{equation}\label{diffeq}
    \rho_0\,\frac{\partial}{\partial t}C_{uu}(k,t) = -\eta_0\,k^2\, C_{uu}(k,t) \, . 
\end{equation}
Combining this with equipartition, $C_{uu}(k_z,0) = \rho_0 k_B T$, the solution to \eq{diffeq} is
\begin{equation}
    C_{uu}(k,t)= \rho_0 k_B T\, e^{-\eta_0 k^2 t / \rho_0} .
    \label{eq:tacf:exp}
\end{equation}
\Fig{fig:tacf:exp} compares this prediction (solid lines) to data obtained at the reference state point $(n_0,T_0)=(1.02,2.58)$ (points) for four wave vectors. The model agrees well with data for the longest wavelength, but is not able to predict the data accurately at shorter wavelengths. This signals a breakdown of classical hydrodynamics. As we show below, this can be remedied by a simple generalization of the Newtonian model {\black to involve the} k-dependent shear viscosity. The value of the shear viscosity $\eta_0$ used in the fit of \fig{fig:tacf:exp} is found in the next section.

\begin{figure*}
\centering
\includegraphics[width=0.7\textwidth]{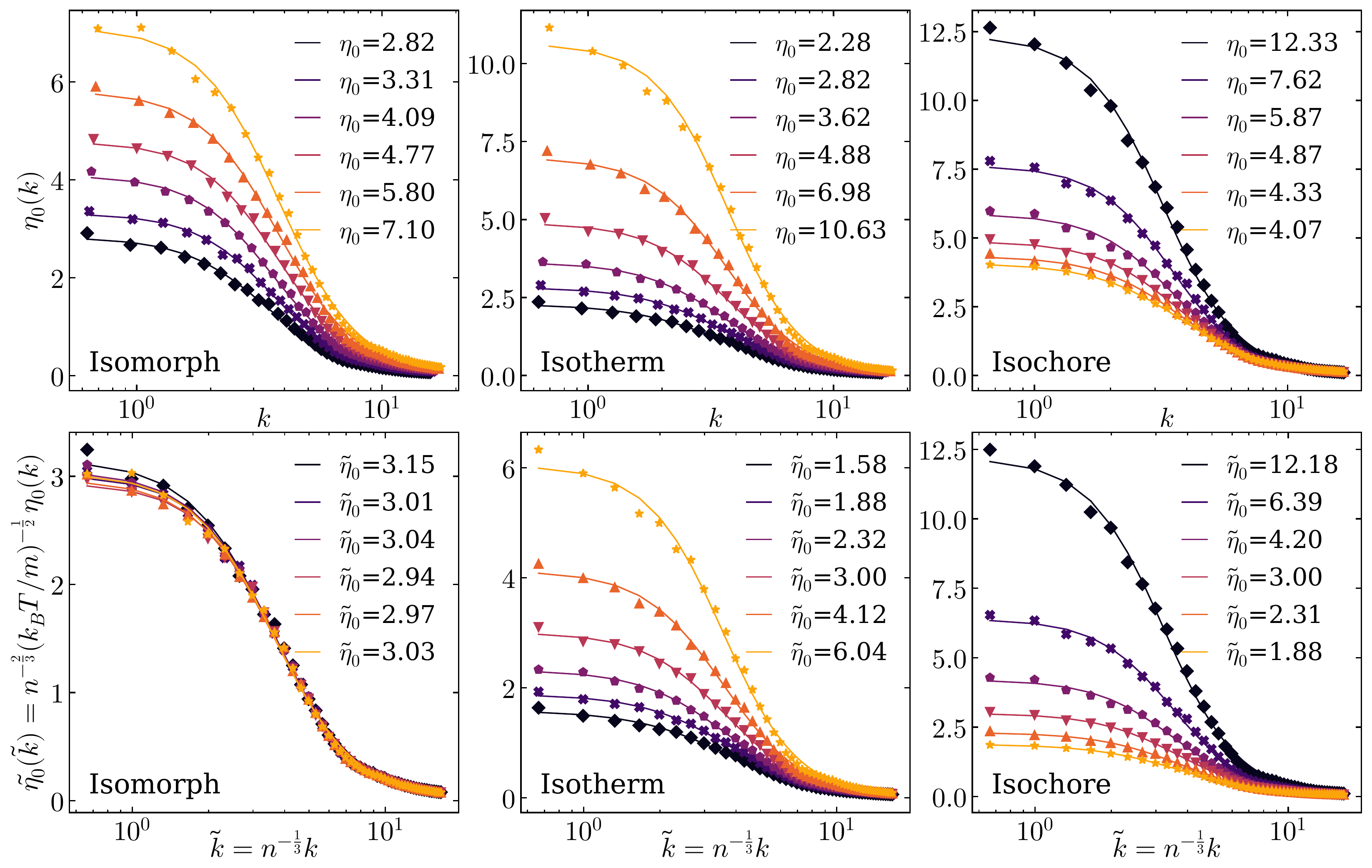}
\caption{The $k$-dependent viscosity $\eta_0(k)$ shown in MD units (top row) and in reduced units (bottom row). The first column is the isomorph, the middle is the isotherm, and the last is the isochore. All data sets are fitted to the expression given in \eq{eq:fit:fct} where the individual $\eta_0(k)$ values are found from simulation data by means of \eq{etanulk}. The $k\to 0$ limit results in a value of $\eta_0$ that is reported for each state point. {\black The fit parameters $\alpha$ and $\beta$ do not vary much, and only one example of their values are thus given. Here for the reference state point $(n=1.02,T=2.58)$: $\alpha=0.03$ and $\beta=2.69$.}}
\label{fig:viscositykernel:1}
\end{figure*}

\subsection{Multiscale viscous response}
In the limit of long times (small frequencies) it is possible to model the TMACF using a wave-vector-dependent (frequency-independent) shear viscosity, $\eta_0(k)$, a quantity that in the $k\rightarrow 0$ limit reduces to the standard macroscopic zero-frequency shear viscosity $\eta_0$. The ``multiscale'' model prediction \cite{Hansen_Dyre_Daivis_Todd_Bruus_2015} is the following generalization of \eq{eq:tacf:exp}
\begin{equation}
    C_{uu}(k,t) = \rho_0 k_B T e^{-\eta_0(k) k^2 t /\rho_0} \,.
\end{equation}
Performing a Fourier-Laplace transform of this we obtain 
\begin{equation}
    \widehat{C}_{uu}(k, \omega) = \frac{ \rho_0k_B T}{\eta_0(k) k^2/\rho_0 + i \omega} \,.
\end{equation}
From this one can calculate $\eta_0(k)$ from simulation data by means of
\begin{equation}\label{etanulk}
    \eta_0(k) = \frac{\rho_0^2 k_BT}{k^2 \widehat{C}_{uu}(k,0)} \,.
\end{equation}
Different expressions for the functional form of $\eta_0(k)$ have been proposed, see, e.g., Refs. \cite{furukawa_nonlocal_2009,martin_transferable_1998,hansen_parameterization_2007}. Here we use the ansatz of Ref. \onlinecite{hansen_parameterization_2007} with $\alpha$ and $\beta$ being fit parameters,
\begin{equation}
    \eta_0(k) = \frac{\eta_0}{1+\alpha k^{\beta}} \,.
    \label{eq:fit:fct}
\end{equation}
In \fig{fig:viscositykernel:1} this expression is fitted to the zero-frequency limit of the Fourier-Laplace transformed data for the wave-vector-dependent TMACF. The top row is in MD units and the bottom row is in reduced units. Notice the different scales of the $y$-axes. When performing a Fourier-Laplace transform on finite data it is important that these are fully relaxed to zero at long times. As mentioned earlier, this is not the case for the smallest wave vector $k\sim 1/3$ (\fig{fig:tacf:MD}). This causes some irregularities for the first data point, which has therefore been removed across all data sets. 

The reduced-unit data for the isomorph in the lower left corner collapse onto a common curve. This shows that not only is the macroscopic shear viscosity isomorph invariant, this quantity is invariant on all length scales. The reduced-unit macroscopic shear viscosity of the LJ system is $\tilde{\eta}_0 = 3.02 \pm 0.07$ for the isomorph in question, which is consistent with values found using other methods \cite{costigliola_communication_2018}.

It is possible to define a characteristic viscous length scale $L$ from the fit parameters in \eq{eq:fit:fct} by means of \cite{puscasu_nonlocal_2010}
\begin{equation}
    L = \alpha^{1/\beta}.
\end{equation}
This length is constant in reduced units along the isomorph and thus provides a novel isomorph characterization in terms of a dimensionless hydrodynamic length. For the isomorph studied here $\widetilde{L}=0.273 \pm 0.005$, which is roughly a quarter of the average nearest-neighbor distance. {The length scale is a measure of when the viscous response becomes non-local, and the classical constitutive relation fails. The important scale is 2*pi*L, such that if the strain rate varies over this length scale the local picture fails.} A study of how $\widetilde{L}$ changes with the isomorph would be very interesting but goes beyond the present work that focuses on a single isomorph.

\subsection{Transverse waves in the multiscale model}
The viscoelastic effects seen in both the shear stress and the transverse momentum time-autocorrelation functions at small wavelengths (large $k$ vectors) can be modelled using a generalized Maxwell viscoelastic model. Recall that the standard Maxwell model is based on the ansatz \cite{phan-thien_understanding_2017,trachenko_collective_2016},
\begin{equation}
    \frac{\partial \delta u_x}{\partial z} = -\frac{1}{\eta_0}\left(1+\tau_M\frac{\partial}{\partial t}\right) P_{zx} \,
\end{equation}
in which $\tau_M$ is the Maxwell relaxation time. In Fourier space this becomes
\begin{equation}
    ik \delta u_x (k,t) = -\frac{1}{\eta_0}\left(1+\tau_M\frac{\partial}{\partial t}\right) P_{zx}(k,t) \, .
\end{equation}
The generalization of this to a $k$-dependent Maxwell relaxation time is
\begin{equation}
    ik \delta u_x (k,t) = -\frac{1}{\eta_0(k)}\left(1+\tau_M(k)\frac{\partial}{\partial t}\right) P_{zx}(k,t) \, .
\end{equation}
Substituting in Eq. (\ref{eq:momentbalance}) and its time derivative one gets 
\begin{equation}
\frac{\partial^2}{\partial t^2}\delta u_x(k,t) + \frac{1}{\tau_M(k)}\frac{\partial}{\partial t}\delta u_x(k,t) + c_T^2(k)k^2 \delta u_x(k,t) =     0 \,.
\end{equation}
We have here introduced the $k$-dependent shear-wave sound velocity given by $c_T^2(k) \equiv \eta_0(k)/(\rho_0\tau_M(k))$. Multiplying by $\delta u_x(-k,0)$ and ensemble averaging we obtain the following differential equation for the TMACF
\begin{equation}
    \frac{\partial^2}{\partial t^2}C_{uu}(k,t) + \frac{1}{\tau_M(k)}\frac{\partial}{\partial t}C_{uu}(k,t) + c_T^2(k)k^2 C_{uu}(k,t) = 0 \, .
\end{equation}
Applying equipartition, $C_{uu}(k,0)= \rho_0 k_B T$, we have whenever $k > 1/(2c_T(k)\tau_M(k))$ (complex eigenvalues) the real and even solution
\begin{equation}
\label{eq:viscwave}
    C_{uu}(k,t)= \rho_0 k_B T e^{-t/2\tau_M(k)}\cos(\omega_T(k) t) \,.
\end{equation}
This represents a damped oscillation for which the characteristic frequency $\omega_T(k)$ is determined by $\omega_T^2(k) = c_T^2k^2 - 1/(4\tau_M^2(k))$. 
Fitting Eq. (\ref{eq:viscwave}) to the $k\sim 3$ isomorph state-point data we find for the reduced shear-wave velocity $\widetilde{c}_T=3.7 \pm 0.1$.

\begin{figure}
\centering
\includegraphics[width=0.4\textwidth]{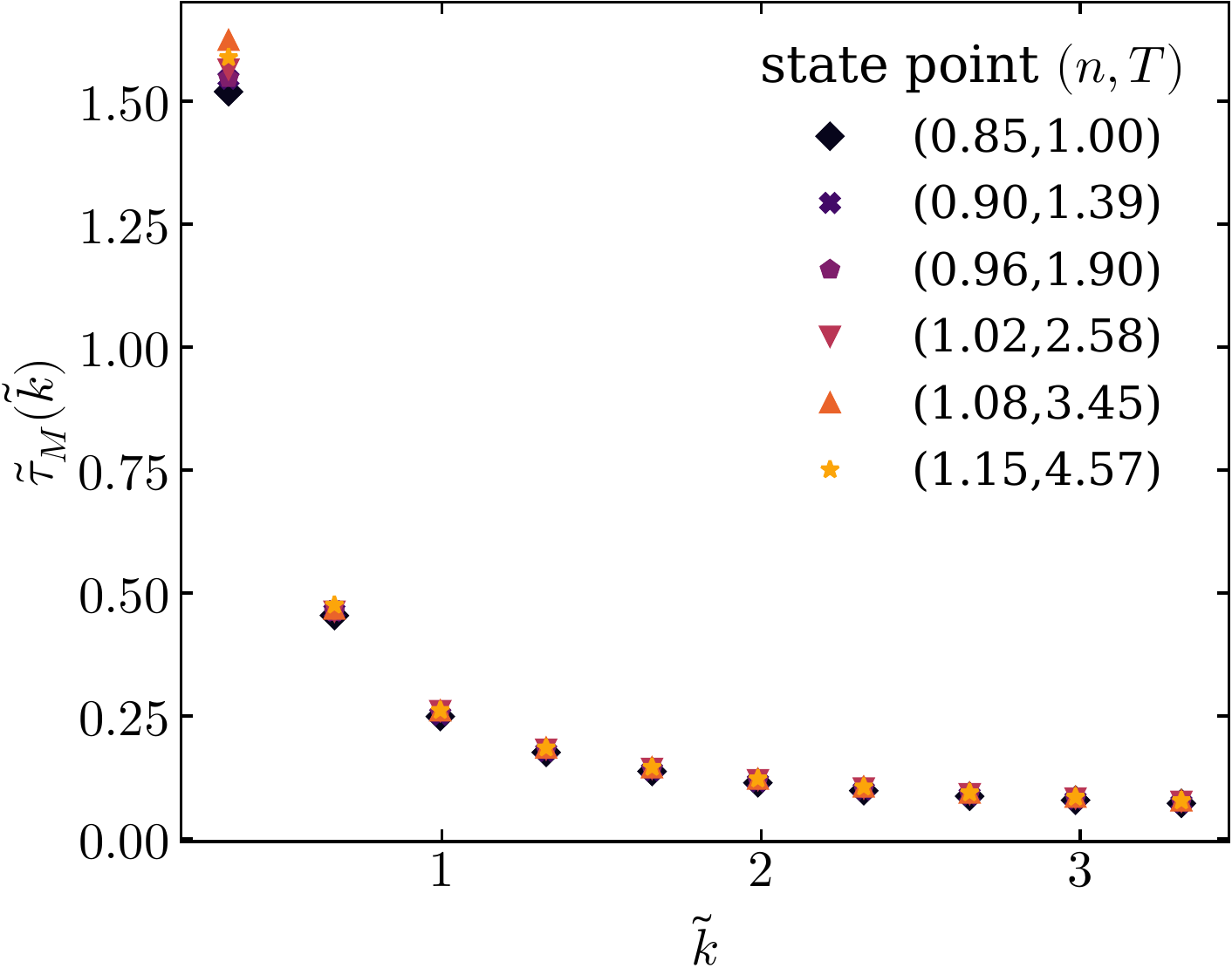}
\vspace*{\floatsep}%
\includegraphics[width=0.4\textwidth]{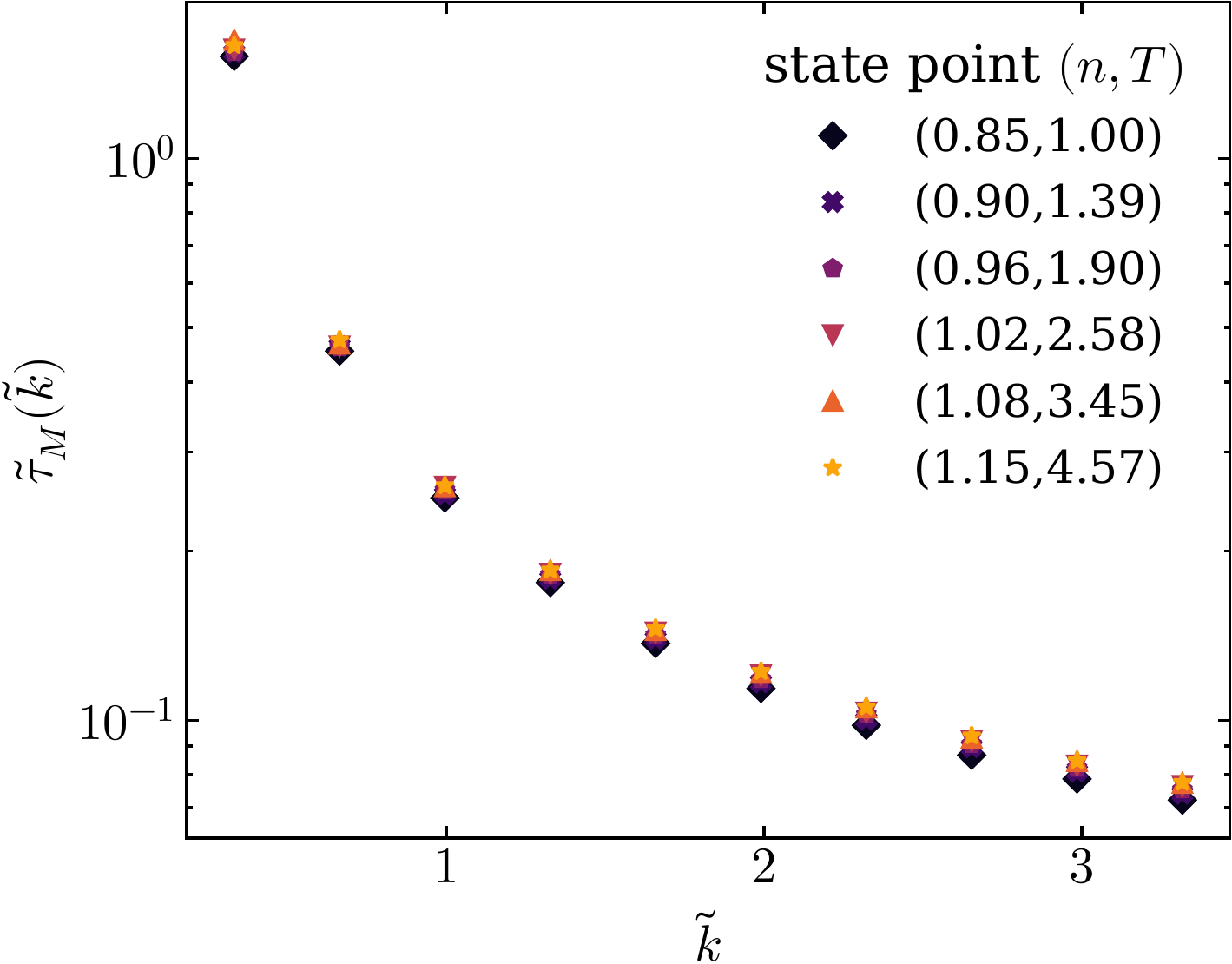}
\caption{The reduced-unit Maxwell relaxation time as a function of the reduced wave vector on linear and logarithmic scale. In the first figure the absolute values are compared and the first point appears to have a larger variance than the rest. In the second figure, however, it is clear that when considering the relative variance, the Maxwell time is approximately isomorph invariant for the range of wave vectors studied.}
\label{fig:maxwell}
\end{figure}

One expects that as the dynamics is isomorph invariant in reduced units, the Maxwell relaxation time is also isomorph invariant in its wave-vector dependence. Figure \ref{fig:maxwell} shows the Maxwell relaxation time as a function of wave vector in the large wave-vector regime (in reduced  units). Since the modulus of rigidity, $G_\infty$, is given by $G_\infty = \eta_0/\tau_M$, we note that this quantity in reduced units is also constant along the isomorph in its $k$ dependence. 

\section{Results for the longitudinal hydrodynamics}
This section presents the simulation data for the longitudinal dynamics, specifically the static and dynamic structure factors.

\subsection{Static structure factor}\label{sec:sk}
\begin{figure*}
	\centering
	\includegraphics[width=0.8\textwidth]{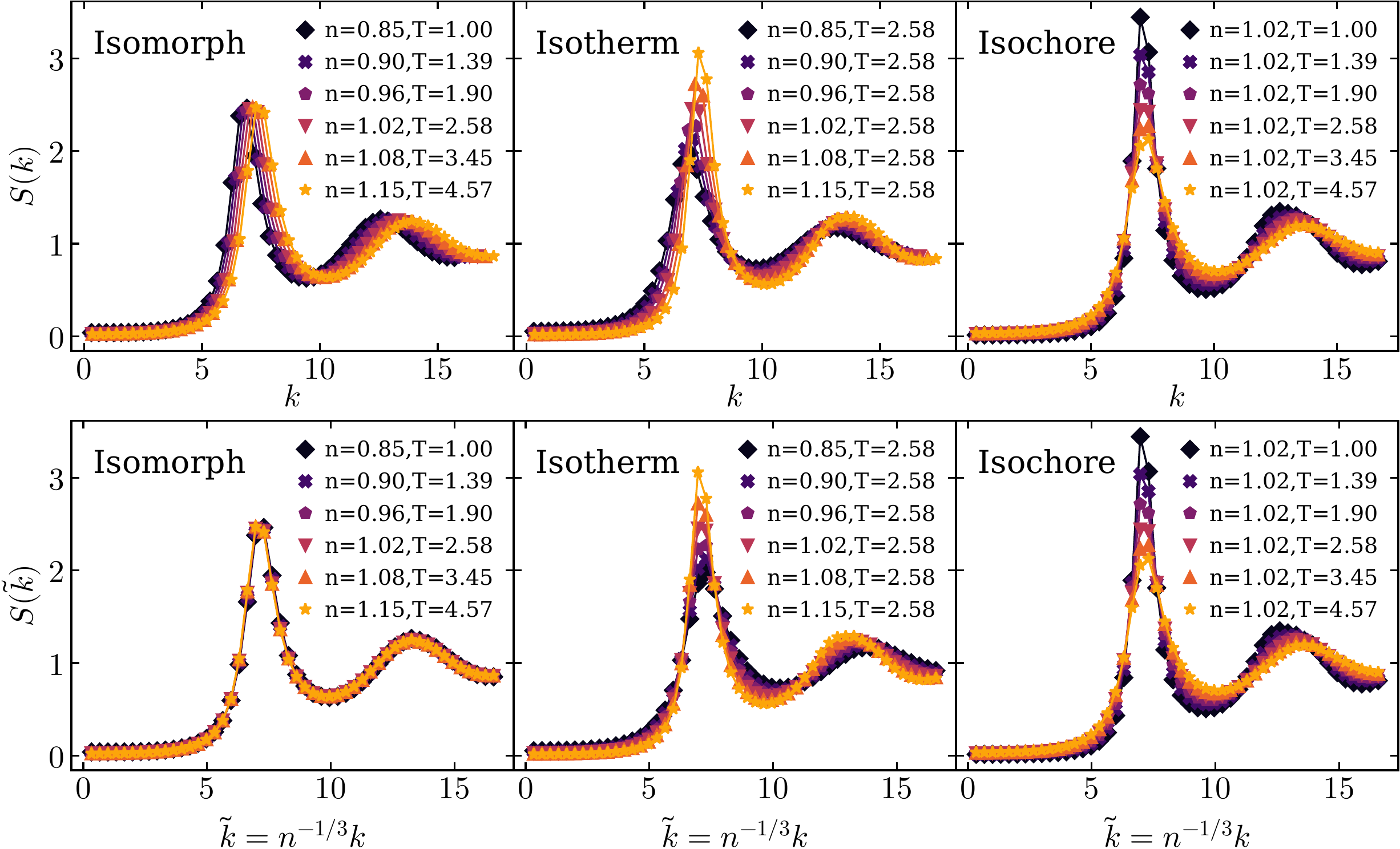}%
	\caption{The static structure factor for state points along the isotherm, isochore, and isomorph. Obtained from the density autocorrelation function (DACF). Upper row: Results in MD units, lower row: Same results in reduced units.}
	\label{fig:Sk}
\end{figure*}

The static structure factor $S(k)$ quantifies the liquid structure. It can be determined in various ways; here we exploit its relation to the DACF following \eq{eq:Fkt:DACF} and (\ref{eq:Sk:Fkt})
\begin{equation}
S(k) = \frac{1}{n} C_{\rho \rho}(k,t=0)/m^2,
\end{equation}
				Another way is to compute the radial pair-distribution function and take the Fourier transform. However, due to finite-size effects, this approach will give small non-physical oscillations in the low $k$-limit, which is one of our areas of interest. Thus, this approach is not pursued further. 

				In \fig{fig:Sk} the static structure factor is shown along the isomorph, isotherm, and isochore, with the top row in MD units and the bottom row in reduced units. The points represent the data from the DACF, and the lines are added as a guide to the eye. From this plot $S(k)$ appears to be isomorph invariant to a good approximation for all $k$.

				For a system with perfect isomorphs (corresponding to $R=1$ for the virial potential-energy correlation coefficient (Eq. (23)) all static and dynamic quantities are isomorph invariant when given in reduced units. This applies only, however (as mentioned in the introduction), for systems with a perfectly Euler homogeneous potential-energy function, i.e., obeying 
				$U(\lambda\vb R)=\lambda^{-n} U(\vb R)$ for some $n$. Approximate isomorph invariance applies much more broadly, for instance for the LJ system in its condensed (liquid and solid) phases. In the more general case, some quantities are more isomorph invariant than others. In particular, quantities like the pressure and the adiabatic or isothermal bulk modulus are generally not isomorph invariant, even in reduced units. This may be understood as follows \cite{Gnan_Schrder_Pedersen_Bailey_Dyre_2009}. For an Euler-homogeneous system, the potential-energy function $U(\vb R)$ can be written $U(\vb R)=h(\rho)\Phi (\vb{\tilde R}) $ for some scaling function $h(\rho)$ and some function of the reduced coordinates $\Phi (\tilde{\vb{R}})$. It is straightforward to show that this implies perfect isomorph invariance of all reduced-unit structure and dynamics. If there are merely strong virial potential-energy correlations, the following identity applies to a good approximation $U(\vb R) = h(\rho)\Phi(\tilde{\vb R})+g(\rho)$ {\red \cite{dyre_isomorphs_2013} }. In this case the reduced-unit structure and dynamics are still approximately isomorph invariant, but the additive factor $g(\rho)$ affects the pressure and its volume derivatives in a way that is unrelated to isomorphs, i.e., in a generally non-invariant way. Thus any quantity, the definition of which in terms of the potential energy involves a perturbation that changes the volume (like pressure and bulk modulus), is generally not isomorph invariant \cite{heyes_transport_2019}.

\begin{figure*}[t]
					\centering
					\includegraphics[width=0.8\textwidth]{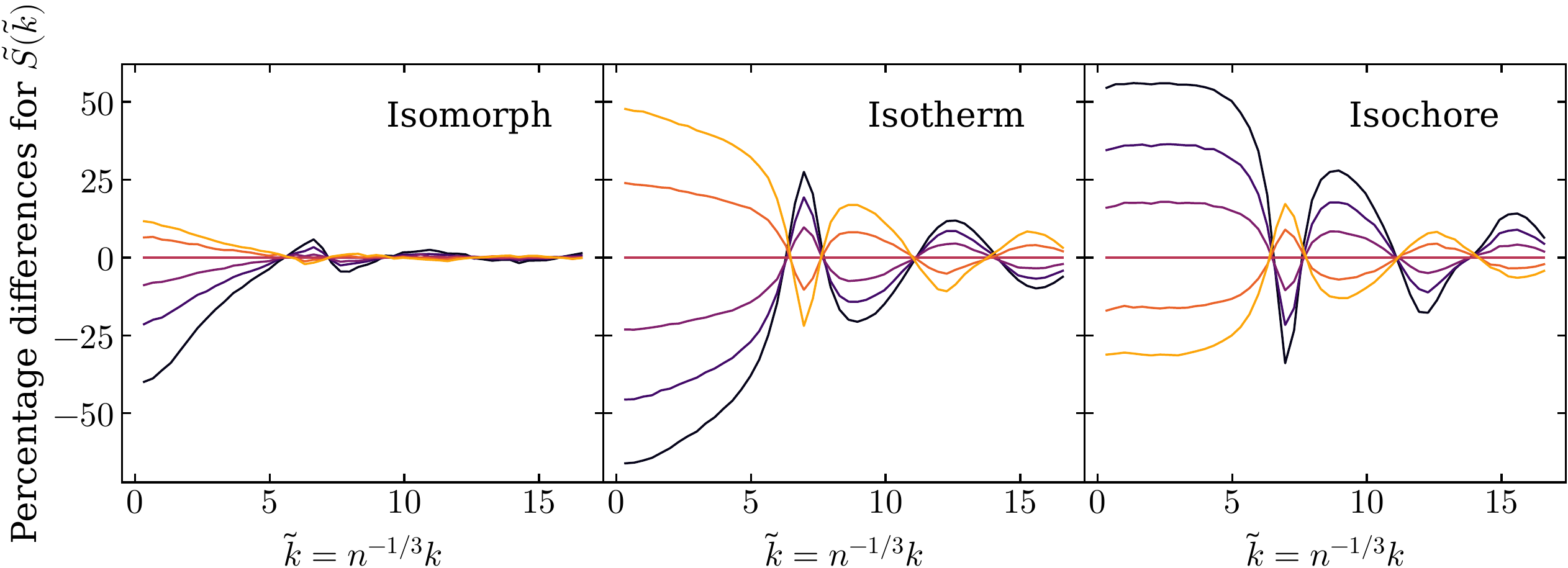}
					\caption{Percentage differences from the reference state point ($n=1.02,T=2.58$) for the static structure factor along the isomorph, isotherm and isochore illustrating that the static structure factor is in fact not invariant for the hydrodynamic limit $k\rightarrow0$ and up till values of $k\sim5$.}
					\label{fig:Sk:pdiff}
				\end{figure*}

				Since the small $k$-limit of the static structure factor is proportional to the isothermal compressibility, this limit is not expected to be isomorph invariant. The percentage differences of the state points from the reference state point $(1.02, 2.58)$ are shown in \fig{fig:Sk:pdiff}, from which it is clear that $S(\vb k)$ is in fact not isomorph invariant near the hydrodynamic limit. The isotherm and isochore data are again shown for comparison.

				\subsection{Dynamic structure factor}
\begin{figure*}
					\centering
					\includegraphics[width=0.8\textwidth]{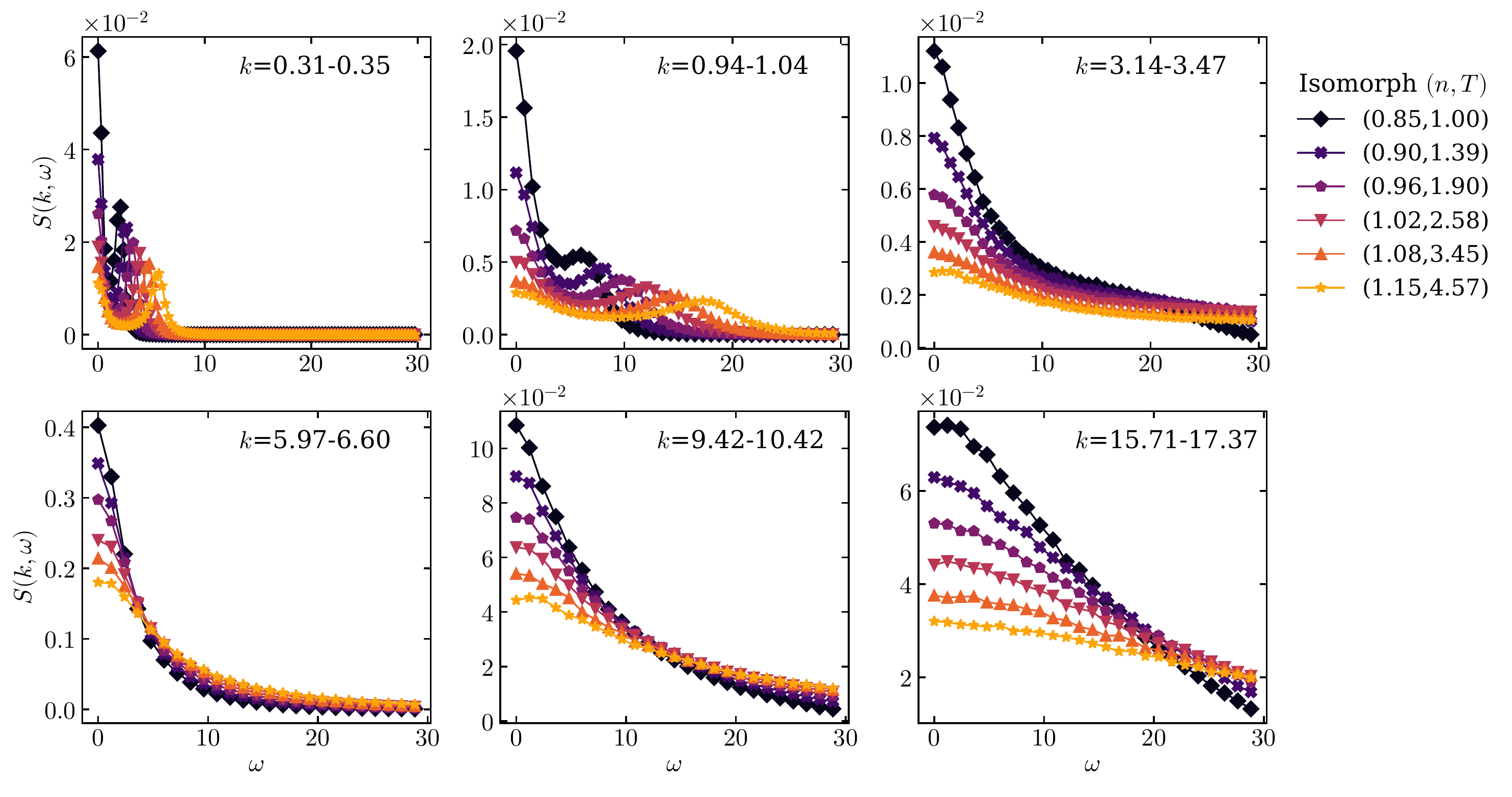}
					\caption{Dynamic structure factor along the isomorph for six different length scales ($k\sim0.3$ to $k\sim16$).}
					\label{fig:Skw:MD:isomorph}
					\vspace*{\floatsep}%
					\includegraphics[width=0.8\textwidth]{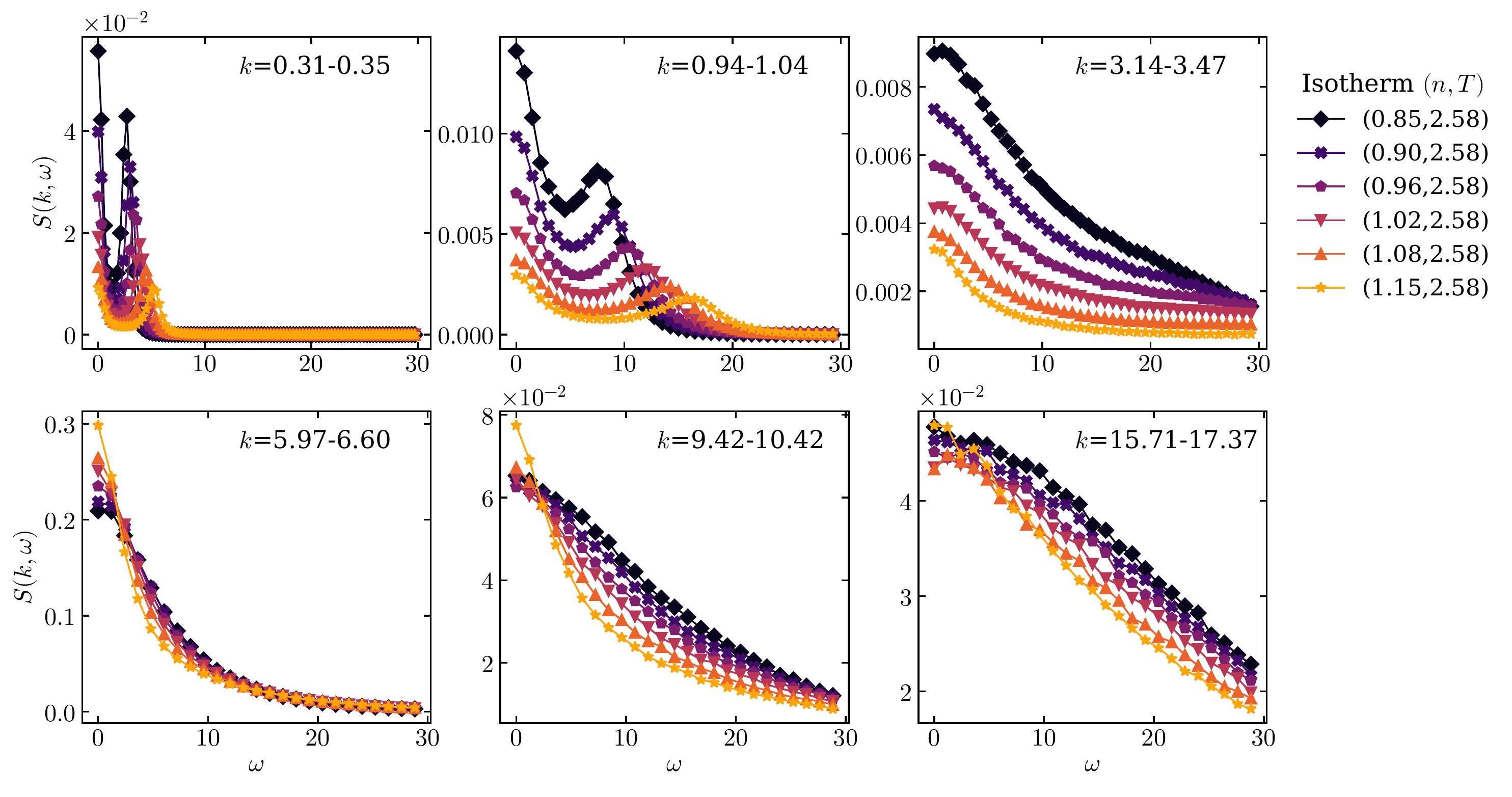}
					\caption{Dynamic structure factor along the isotherm for six different length scales ($k\sim0.3$ to $k\sim16$).}
					\label{fig:Skw:MD:isotherm}
					\vspace*{\floatsep}
					\includegraphics[width=0.8\textwidth]{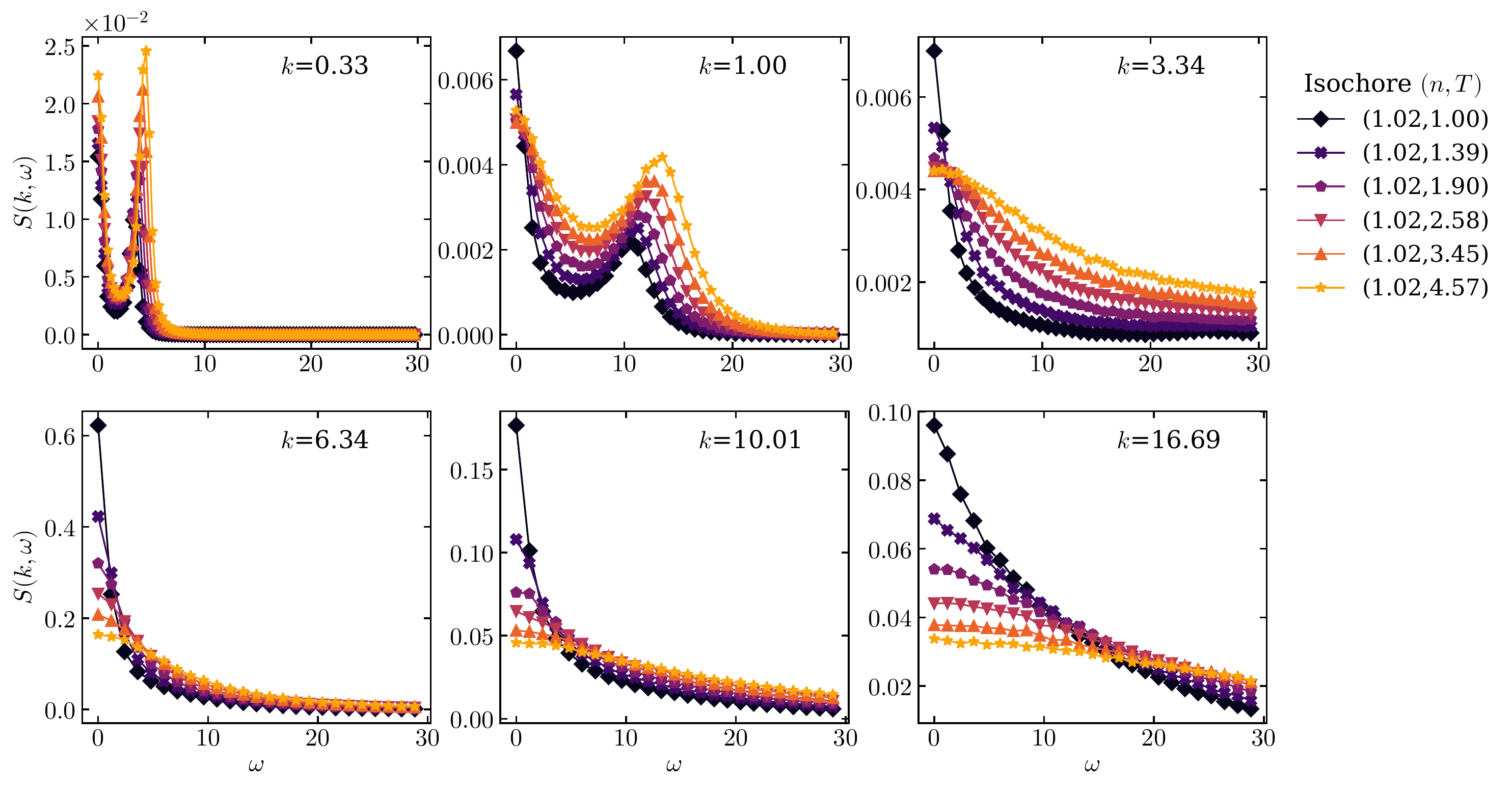}
					\caption{Dynamic structure factor along the isochore for six different length scales ($k\sim0.3$ to $k\sim16$).}
					\label{fig:Skw:MD:isochore}
				\end{figure*}

				The longitudinal dynamics is investigated below by the dynamic structure factor. Note that, while other correlation functions can be studied, this contains all relevant information of the physical processes under interest. Data in MD units for the isomorph, isotherm and isochore can be seen in \fig{fig:Skw:MD:isomorph}, \fig{fig:Skw:MD:isotherm}, and \fig{fig:Skw:MD:isochore}, respectively. The six panels each show $S(k,\omega)$ at approximately the same length scale for each of the six state points, starting with the longest length scale ($k\sim0.3$) in the upper left corner and ending with the shortest ($k\sim16$) in the lower right corner. The six length scales shown here are sought to represent the way $S(k,\omega)$ changes qualitatively with decreasing length. Notice the change of scale on the vertical axes. The Rayleigh peak at the origin corresponds to the thermal diffusion process and the Brillouin peak (second peak) to the adiabatic propagating sound wave. As we go to shorter length scales (higher $k$), the Brillouin peak attenuates and broadens, and when reaching the inter-particle distance  at around $k\sim 2\pi$, the peak seems to vanish or be completely absorbed in the Rayleigh peak. This feature is consistent across the isomorph, isotherm, and isochore. 

\begin{figure*}
					\centering
					\includegraphics[width=0.76\textwidth]{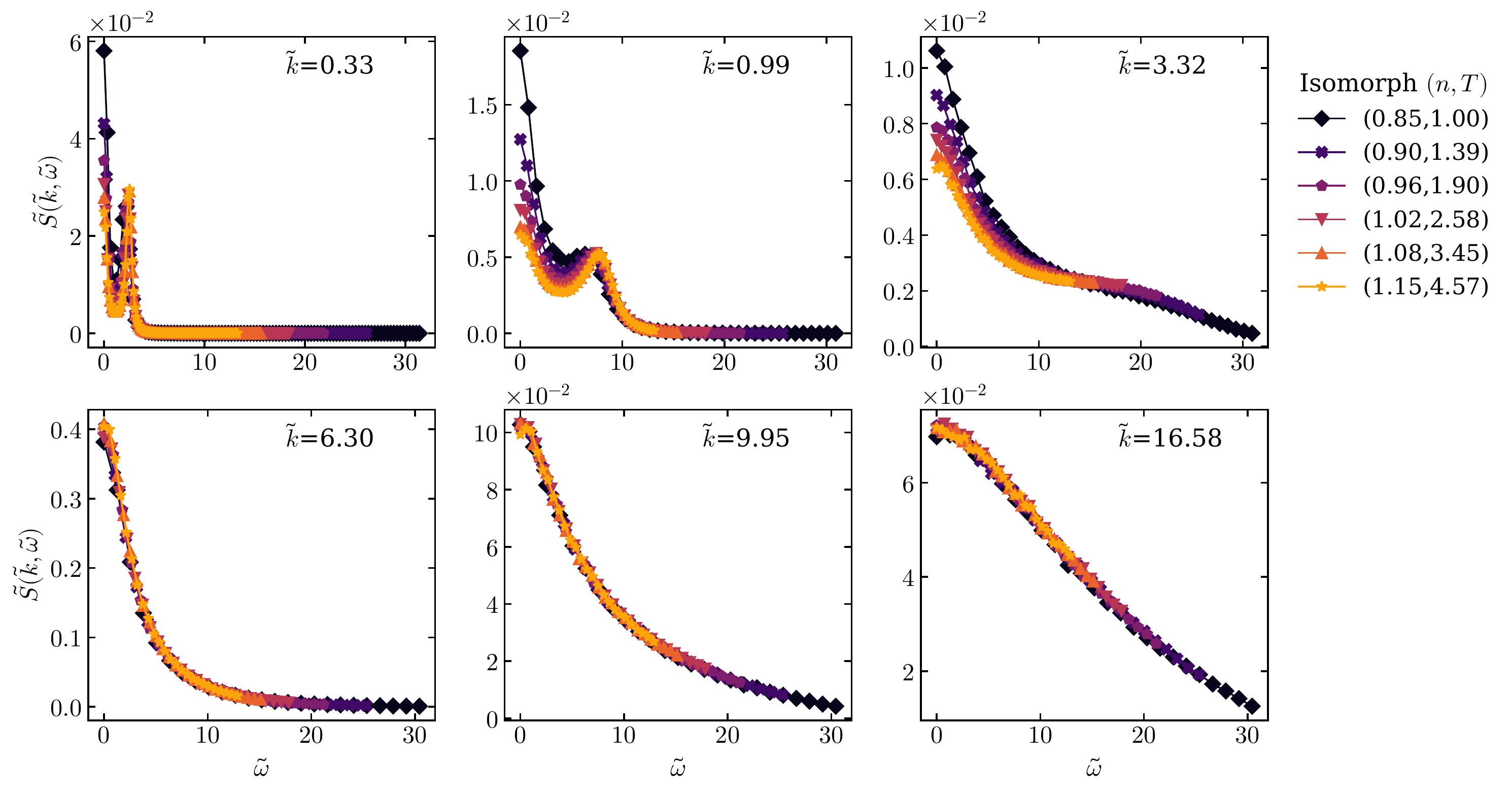}
					\caption{Dynamic structure factor along the isomorph for six different length scales with $\tilde S (\tilde k, \tilde \omega) = n^{1/3} (k_B T/m)^{1/2} \, S(k,\omega)$ and $\tilde \omega = n^{-1/3}(k_B T/m)^{-1/2} \omega$.}
					\label{fig:Skw:MR:isomorph}
					\vspace*{\floatsep}%
					\includegraphics[width=0.76\textwidth]{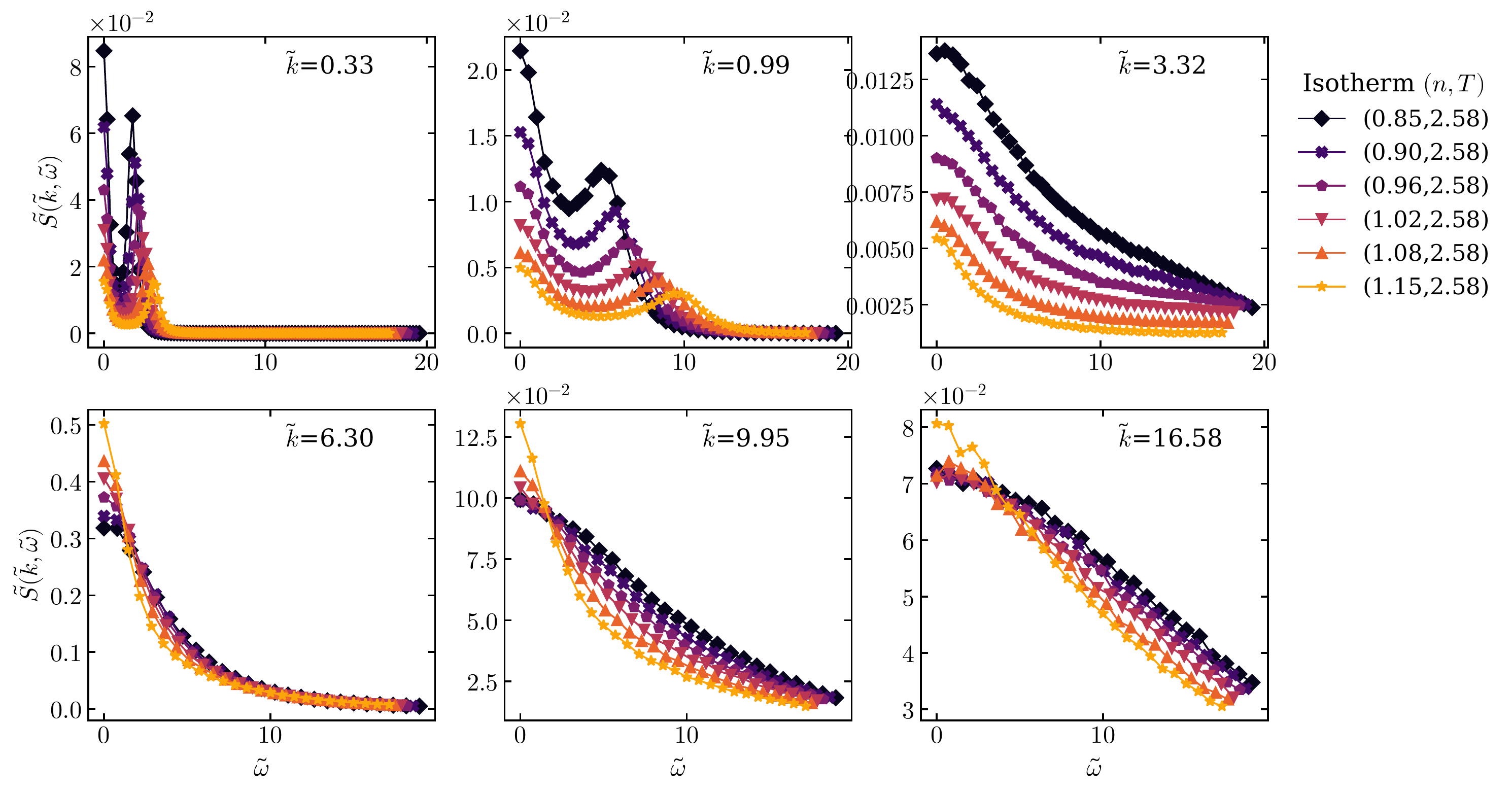}
					\caption{Dynamic structure factor along the isotherm for six different length scales with $\tilde S (\tilde k, \tilde \omega) = n^{1/3} (k_B T/m)^{1/2} \, S(k,\omega)$ and $\tilde \omega = n^{-1/3}(k_B T/m)^{-1/2} \omega$.}
					\label{fig:Skw:MR:isotherm}
					\vspace*{\floatsep}
					\includegraphics[width=0.76\textwidth]{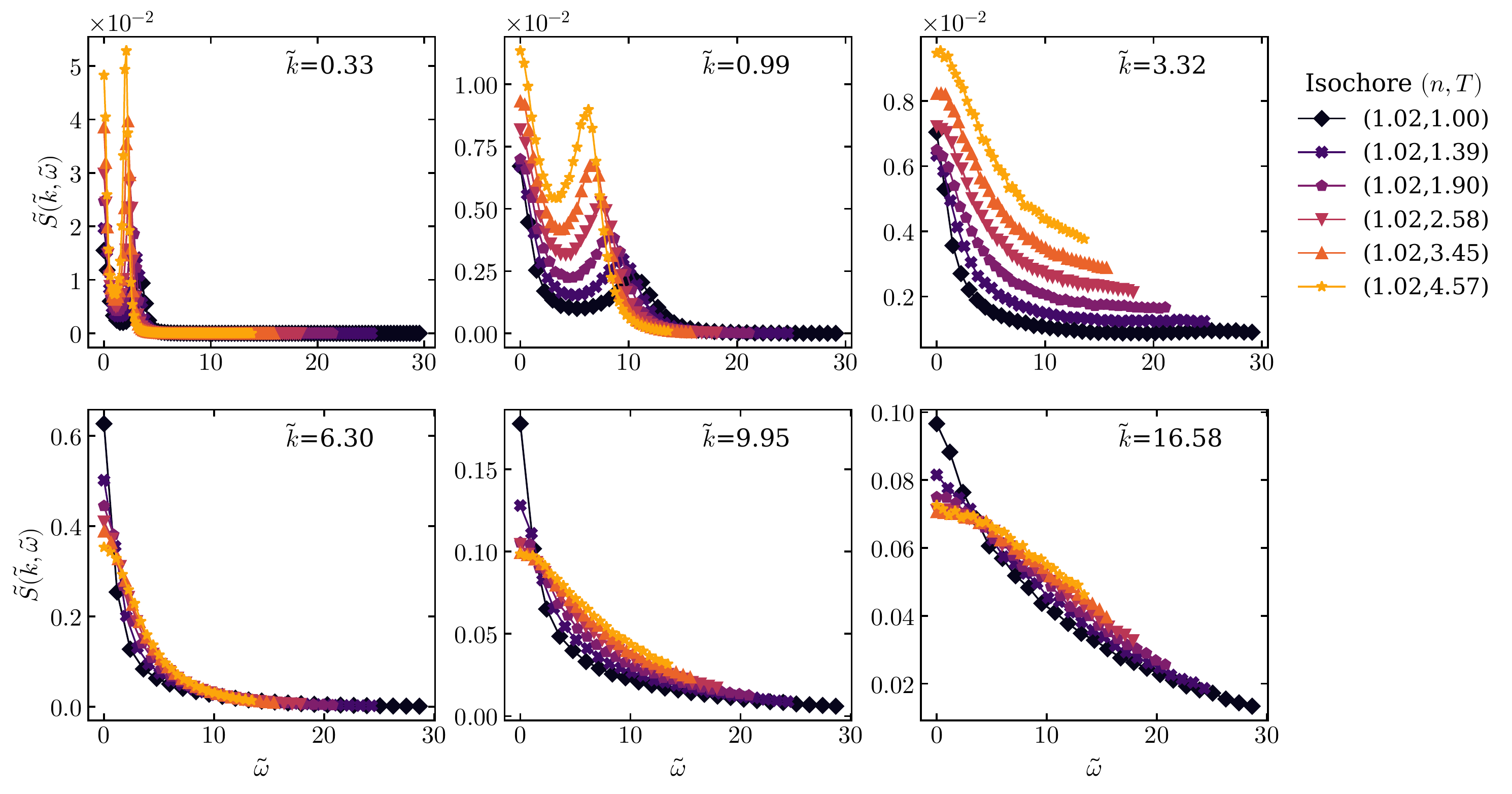}
					\caption{Dynamic structure factor along the isochore for six different length scales with $\tilde S (\tilde k, \tilde \omega) = n^{1/3} (k_B T/m)^{1/2} \, S(k,\omega)$ and $\tilde \omega = n^{-1/3}(k_B T/m)^{-1/2} \omega$.}
					\label{fig:Skw:MR:isochore}
				\end{figure*}

				The reduced-unit dynamic structure factor $\tilde S(\tilde k, \tilde \omega)$ is shown in \fig{fig:Skw:MR:isomorph} (isomorph), \fig{fig:Skw:MR:isotherm} (isotherm), and \fig{fig:Skw:MR:isochore} (isochore). No collapse of curves is observed for the isotherm and isochore, while for the isomorph, data for length scales around $k\sim 2 \pi$ and shorter (higher $k$) collapse. Such a collapse does not happen for the isotherm and isochore and is thus not a trivial consequence of the very short distances. For some reason the short-length dynamics are isomorph invariant whereas the dynamics at length scales larger than the inter-particle distance are not. In the following section, the non-collapsing region is  studied in more details.

				\section{Analysis of the longitudinal hydrodynamics}
				In this section we utilize a hydrodynamic model for the mass density auto-correlation function in order to gain a better understanding of the large length scales at which the reduced-unit longitudinal hydrodynamic data vary along the isomorph.

				\subsection{The hydrodynamic model}
				The following classical hydrodynamic model, found in textbooks such as Hansen and McDonald \cite{hansen_mcdonald_2013}, can be used to fit the DACF data in the hydrodynamic limit $k\rightarrow0$. The fit parameters are the ratio between heat capacities $\gamma_c=c_P/c_V$, the thermal diffusion coefficient $D_T$, the sound attenuation $\Gamma$, and the adiabatic sound velocity $c_s$
				\begin{align}
				C_{\rho\rho}( k, t) = \frac{1}{\gamma_c} \left[ (\gamma_c-1) e^{-D_T k^2 t} + e^{-\Gamma k^2 t} \cos (c_s k t) \right].
				\label{eq:clasHDmodel}
				\end{align}
				The first term models the thermal diffusion (the Rayleigh process), and the second term models the adiabatic sound waves (the Brillouin process). The specific heat ratio $\gamma_c$ provides a measure of the ratio between the two processes.

				Heuristically we let the parameters depend on $k$, i.e., $\gamma_c\rightarrow\gamma_c(k)$, $D_T \rightarrow D_T(k)$, etc., as a way of generalizing \eq{eq:clasHDmodel}
				\begin{multline}
				C_{\rho\rho}( k, t) = \frac{\gamma_c( k)-1}{\gamma_c(k)} \exp[{-D_T( k) k^2 t}] +\\
				\frac{1}{\gamma_c( k)} \exp[{-\Gamma(k) k^2 t}] \cos (c_s( k) k t).
				\label{eq:genHDmodel}
				\end{multline}
				The derivaton of \eq{eq:clasHDmodel} with $k$-space parameters can be shown to give \eq{eq:genHDmodel}, so here we will just refer to the soon-to-be-published book by Hansen \cite{hansenNEWbook}. Now, with everything being $k$-dependent, the physical interpretation of $\gamma_c(k)$ as being the ratio of the heat capacities necessarily depends on the heat capacities being $k$-dependent. A similar argument applies to the thermal diffusion, which depends on the heat conductivity and the heat capacity at constant pressure $D_T(k)=\lambda(k)/\rho_0 c_P(k)$, as well as to the sound attenuation $\Gamma(k)$ and speed of sound $c_s(k)$.

\begin{figure*}
					\centering
					\includegraphics[width=0.75\textwidth]{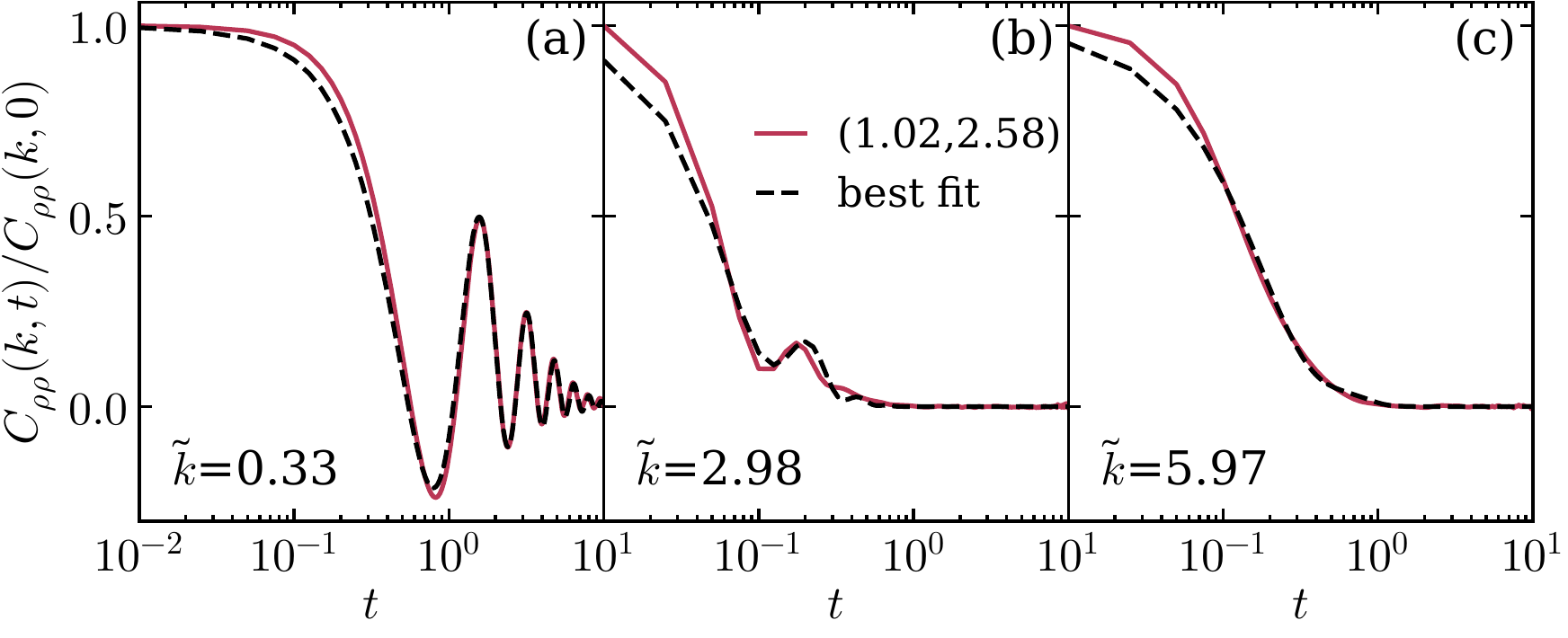} %evt brug spline fct to make the curves smoother
					\caption{Three representative examples of fitting the generalized hydrodynamic expression for the DACF in \eq{eq:genHDmodel}. Data are from the reference state point $(1.02,2.58)$ for $\tilde{k}<2\pi$. It is clear from these figures that the behaviour of the DACF changes significantly, with (a) showing clear oscillations and negative correlations, (b) showing only a tiny oscillation and positive correlations, and finally in (c) all oscillations have been attenuated, leaving only a decaying exponential. Notice that the data are normalized.}
					\label{fig:lmfit}
				\end{figure*}

				Three representative examples of fitting the generalized model to the DACF data for different length scales can be seen in \fig{fig:lmfit} in which all four $k$-dependent parameters are used as fit parameters. The data change significantly over the range of length scales studied, with dampened oscillations for the longest length scales and a simple exponential decay for the shortest length scale. This reproduces the loss of features observed for $S(k,\omega)$ in \fig{fig:Skw:MD:isomorph} for decreasing wave lengths. 

\begin{figure*}
					\centering
					\includegraphics[width=\textwidth]{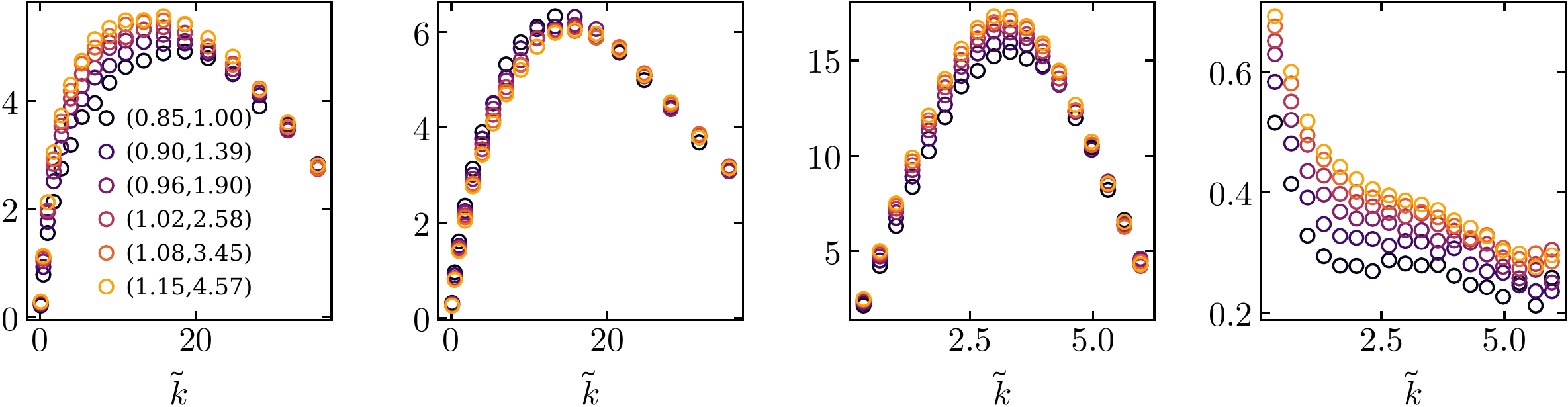}
					\caption{Dispersion curves along the isomorph for the three fit parameters $\tilde D_T(k) \tilde k^2$, $\tilde \Gamma(k) \tilde k^2$, and $\tilde c_s(k) \tilde k$, as well as for $\tilde 1/ \gamma_c(k)$.}
					\label{fig:disp:IM:onego:lin}
				\end{figure*}

				The fit parameters can be seen in \fig{fig:disp:IM:onego:lin}, as a function of $\tilde k$ and $\tilde k^2$,
				respectively. As expected \cite{hansen_mcdonald_2013} the three dispersion plots for the frequencies $\widetilde{D_Tk}^2$, $\widetilde{\Gamma k}^2$, and $\widetilde{c_s k}$ decrease toward zero for higher $k$. {\black The  generalized ratio of heat capacities $1/\gamma_c(k)$} decreases more or less smoothly for higher $k$, with a sharper decrease in the beginning and then reaching more toward a plateau for higher $k$. {\black The length scale where a plateau is reached depends on the state point.} We see that the Brillouin process dominates for small $k$, with the Rayleigh process taking over for higher $k$. This is again in agreement with what we saw for $S(k,\omega)$ in \fig{fig:Skw:MD:isomorph}, where the sound waves disappear for higher $k$.

				\subsection{The classical hydrodynamic limit}
\begin{figure*}[t]
					\centering
					\includegraphics[width=0.8\textwidth]{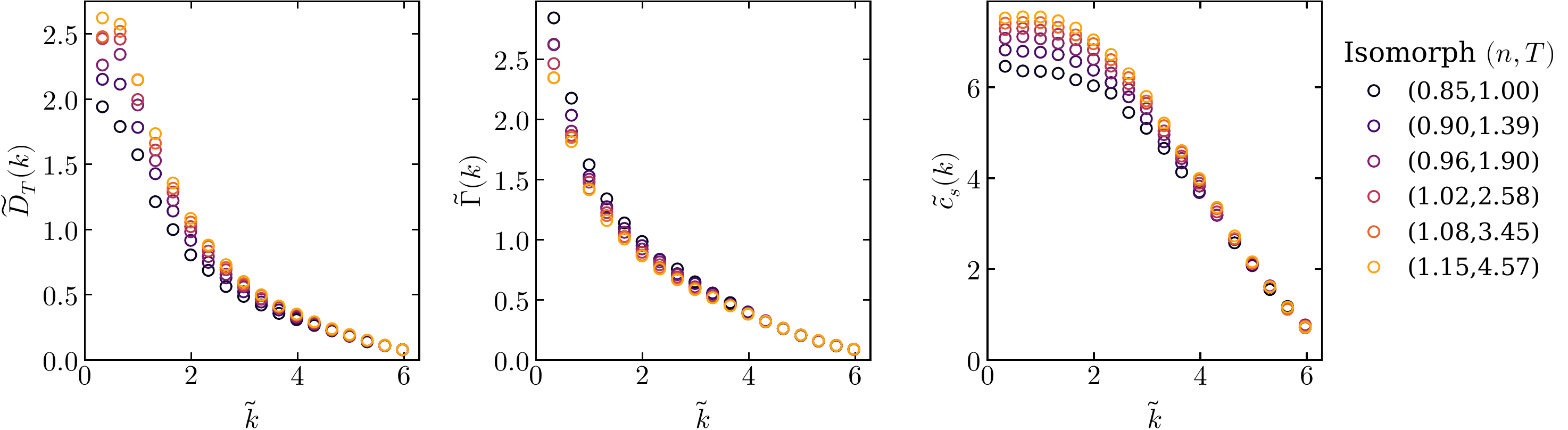}
					\caption{The three fit parameters plotted against $\tilde k$. If the longitudinal dynamics of the system had reached the classical hydrodynamic limit for $\tilde k\rightarrow0$, all three quantities would reach a plateau for $\tilde k\rightarrow0$. We see instead a non-isomorph-invariant plateau for $\tilde c_s$, an indication of a plateau for $\tilde D_T$, and no plateau for $\tilde \Gamma$ at all. In order to reach the classical limit for $\tilde \Gamma$ as well, the system size would have to be increased.}
					\label{fig:HDlim:lin}
				\end{figure*}

				In the classical hydrodynamic limit $ k \rightarrow 0$, the macroscopic quantities $D_T$, $\Gamma$, and $c_s$
				are independent of $ k$. Thus a plot of the generalized quantities against $k$ should reach a plateau in this limit. For the shear viscosity in \fig{fig:viscositykernel:1}, a clear plateau is observed, indicating that the transverse dynamics of our system has reached the classical limit. {\black Data for $\tilde D_T(\tilde k)$, $\tilde \Gamma(\tilde k)$, and $\tilde c_s(\tilde k)$ plotted against $\tilde k$, i.e., in reduced units, are shown in \fig{fig:HDlim:lin}. A plateau is not present for all quantities, indicating that the thermal diffusion coefficient, the sound attenuation, and the longitudinal speed of sound reach the classical limit at different length scales.} The speed of sound clearly shows a plateau and the diffusion coefficient seems to be on the edge of one, whereas the sound attenuation coefficient shows no signs of reaching a plateau at all. 
				{Classical hydrodynamics correctly predicts the transverse dynamics on length scales where the predictions for the longitudinal dynamics fail. This means that it is the simple linear diffusion of momentum which is the governing process for the transverse dynamics, but that the underlying processes for the longitudinal dynamics, that is, the relaxation of local density and temperature fluctuations, is not correctly accounted for by the theory. Also see \cite{JSHmolsim}.}

				\subsection{{\black Scaling of the} $\tilde k$-region below $2\pi$}
				{\black Staying with \fig{fig:HDlim:lin}, but} turning our attention to the full non-collapsing region of $\tilde k$-vectors, a general trend is observed where the collapse gradually gets better for increasing $\tilde k$-vectors for all three wave-dependent parameters. 

\begin{figure*}[t]
					\centering
					\includegraphics[width=0.8\textwidth]{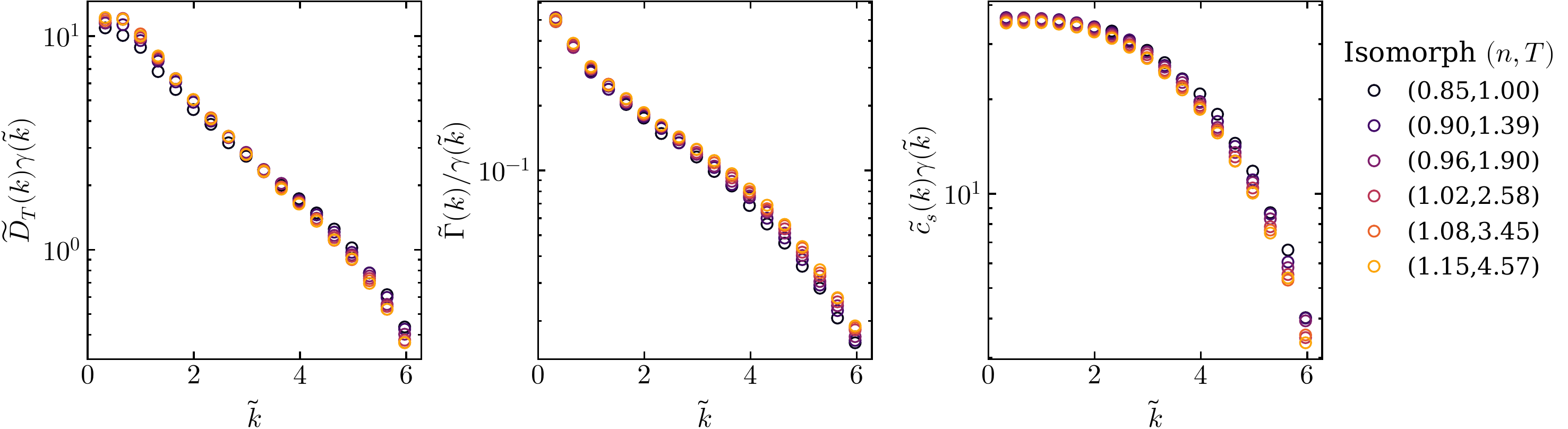}
					\caption{The three fit parameters in reduced units can be scaled with $\gamma$, to make the small $\tilde k$-region data collapse onto a master curve. The reason for this is not understood.}
					\label{fig:HDlim:log:gam}
				\end{figure*}

\begin{figure*}[t]
					\centering
					\includegraphics[height=0.33\textwidth]{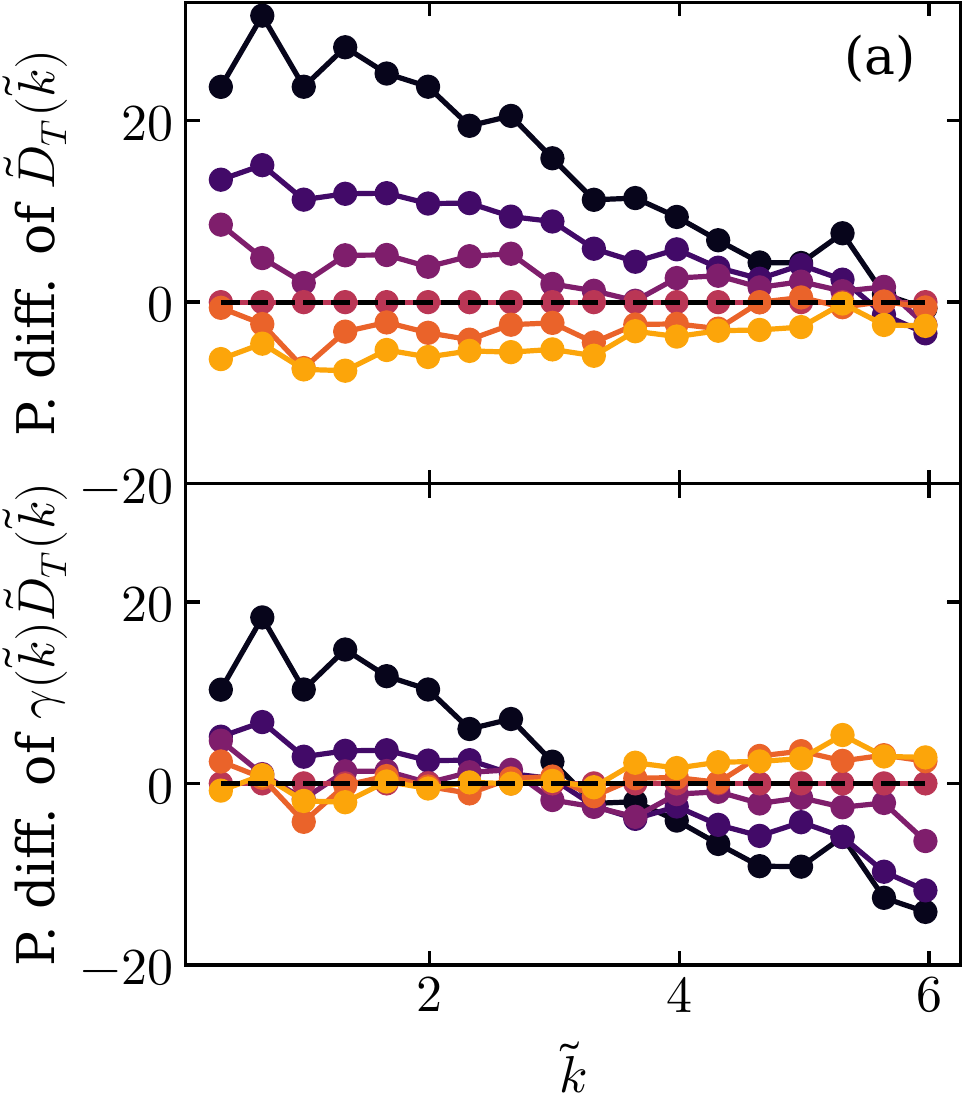}
					\hspace*{\floatsep}%
					\includegraphics[height=0.33\textwidth]{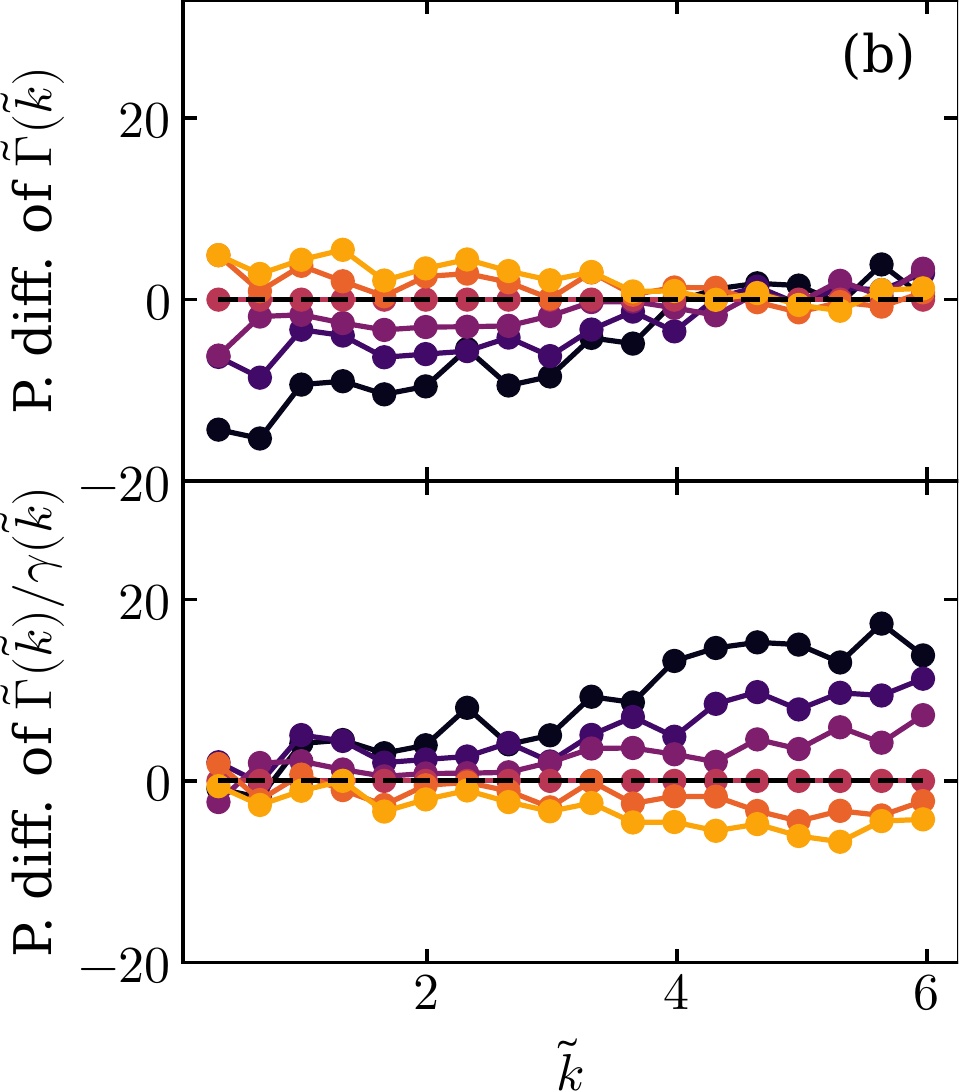}
					\hspace*{\floatsep}%
					\includegraphics[height=0.33\textwidth]{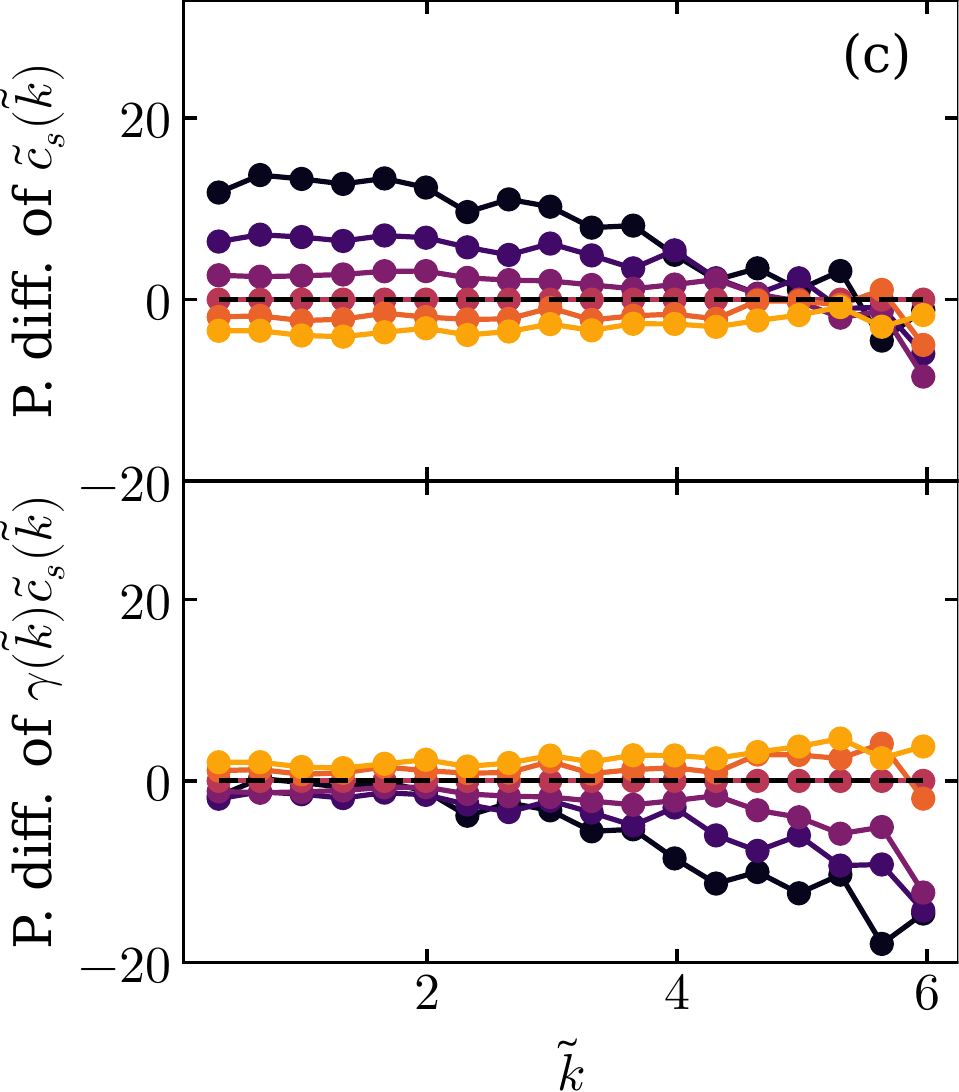}
					\caption{The percentage differences of the fit parameters shown before (upper figures) and after (lower figures) scaling with $\gamma$. Subfigure (a) shows $\tilde D_T(\tilde k)$, (b) shows $\tilde \Gamma(\tilde k)$, and (c) shows $\tilde c_s(\tilde k)$. The percentage differences are taken with respect to the reference state point $(n,T)=(1.02,2.58)$. The coloring is consistent with all other figures.}
						\label{fig:Pdiff:all}
				\end{figure*}

				{\black  The collapse improves with empirical scaling by $\gamma$, one of the fundamental quantities in isomorph theory defined in \eq{eq:gamma} (not to be confused with the generalized ratio of heat capacities $\gamma_c(k)$). $\gamma$ is the state-point-dependent (dimensionless) density-scaling exponent, which intriguingly offers the multiplicative factor needed to shift the collapse in the higher-$k$ region to the low-$k$ region for all three generalized transport coefficients. In \fig{fig:HDlim:log:gam}, the data from \fig{fig:HDlim:lin} are replotted in a log-plot, this time scaled by $\gamma$. In order to quantify the shifting of invariance, the percentage differences between the fit parameter data for the six state points are shown in \fig{fig:Pdiff:all} comparing scaling with $\gamma$ to no scaling.} The adiabatic speed of sound in particular improves its collapse. {\black The adiabatic or longitudinal speed of sound $c_s$ depends on the longitudinal viscosity, which in turn depends on both the bulk and shear viscosities as $\eta_{l}=\eta_b + 4/3\eta_0$. The reduced shear viscosity has been shown to be isomorph invariant in \cite{Costigliola_Schrder_Dyre_2016}. In agreement with that, we have shown that the k-dependent shear viscosity is invariant for a broad range of length scales. In contrast, the bulk viscosity has been shown to not be isomorph invariant \cite{heyes_transport_2019}. Thus the longitudinal viscosity would be expected to carry this non-invariance on to the speed of sound, as is also observed. It would be interesting to study how the non-invariance of the bulk viscosity behaves on different length scales, and whether it behaves similar to the speed of sound. Additionally, it would be interesting to see if the non-invariance of the bulk viscosity for long wavelengths could be scaled into invariance by $\gamma$ as well.}

				This empirical scaling was inspired by ref. \cite{bell_excess-entropy_2020}, which used $\gamma$ as a correction parameter when comparing different systems to obtain a better isomorph collapse. It is still not understood how this may relate to the present observation, where scaling with $\gamma$ is used to improve a collapse along a specific isomorph. 

				\section{Summary}
				The transverse collective dynamics of the LJ liquid system has been studied through the time dependence of both the shear-stress autocorrelation function and the transverse momentum autocorrelation function. These have both been shown to be invariant to a good approximation along the isomorph when given in reduced units. The same invariance does not apply along the isotherm or the isochore. Our results constitute the first systematic study and confirmation of isomorph properties involving generalized hydrodynamics. For the longitudinal generalized hydrodynamics the situation is more complex: these dynamics are isomorph invariant to a good approximation at large and intermediate wave vectors, but not at small wave vectors, i.e., in the hydrodynamic limit. We report an empirical scaling making the dynamics at small wave vectors invariant, though at the cost of loosing some invariance at intermediate and large wave vectors.

				A non-standard approach was used to define the wave-dependent shear stress, which approaches the standard Green-Kubo expression in the limit $k\rightarrow0$. Using a hydrodynamic model the shear viscosity was determined across a broad spectra of wave vectors, and found to be isomorph invariant for all. A novel characterization of isomorphs through a dimensionless hydrodynamic length has also been proposed. 

				\begin{acknowledgments}
				The authors would like to thank Lorenzo Costigliola for suggesting the scaling with $\gamma$, as well as checking our value of macroscopic shear viscosity $\eta_0$ with existing methods. This work was supported by the VILLUM Foundation's \textit{Matter} grant (No. 16515).
				\end{acknowledgments}


\begin{thebibliography}{45}
\providecommand{\natexlab}[1]{#1}
\providecommand{\url}[1]{\texttt{#1}}
\expandafter\ifx\csname urlstyle\endcsname\relax
\providecommand{\doi}[1]{doi: #1}\else
\providecommand{\doi}{doi: \begingroup \urlstyle{rm}\Url}\fi

\bibitem[Landau and Lifshitz(1959)]{Landau_Lifshitz_1959}
L.~D. Landau and E.~M. Lifshitz.
\newblock \emph{Fluid Mechanics}, volume~6 of \emph{Course of Theoretical
Physics}.
\newblock Pergamon Press Ltd, 1959.

\bibitem[Batchelor(1967)]{batchelor_introduction_1967}
G.~K. Batchelor.
\newblock \emph{An introduction to fluid dynamics}.
\newblock Cambridge University Press, 1967.

\bibitem[Groot and Mazur(1984)]{Groot_Mazur_1984}
S.~R.~de Groot and P.~Mazur.
\newblock \emph{Non-equilibrium thermodynamics}.
\newblock Dover Publications, dover ed edition, 1984.
\newblock ISBN 9780486647418.

\bibitem[Boon and Yip(1991)]{Boon_Yip_1991}
J.~P. Boon and S.~Yip.
\newblock \emph{Molecular hydrodynamics}.
\newblock Dover Publications, 1991.
\newblock ISBN 9780486669496.

\bibitem[Alley and Alder(1983)]{AlleyAlder1983}
W.~E. Alley and B.~J. Alder.
\newblock Generalized transport coefficients for hard spheres.
\newblock \emph{Physical Review A}, 27\penalty0 (6):\penalty0 3158--3173, Jun
1983.
\newblock \doi{10.1103/PhysRevA.27.3158}.
\newblock URL \url{https://link.aps.org/doi/10.1103/PhysRevA.27.3158}.

\bibitem[Bailey et~al.(2008{\natexlab{a}})Bailey, Pedersen, Gnan, Schr{\o}der,
and Dyre]{Bailey_Pedersen_Gnan_Schrder_Dyre_2008a}
N.~P. Bailey, U.~R. Pedersen, N.~Gnan, T.~B. Schr{\o}der, and J.~C. Dyre.
\newblock Pressure-energy correlations in liquids. {II}. {A}nalysis and
consequences.
\newblock \emph{The Journal of Chemical Physics}, 129\penalty0 (18):\penalty0
184508, Nov 2008{\natexlab{a}}.
\newblock \doi{10.1063/1.2982249}.
\newblock URL \url{https://aip.scitation.org/doi/full/10.1063/1.2982249}.

\bibitem[Bailey et~al.(2008{\natexlab{b}})Bailey, Pedersen, Gnan, Schr{\o}der,
and Dyre]{Bailey_Pedersen_Gnan_Schrder_Dyre_2008b}
N.~P. Bailey, U.~R. Pedersen, N.~Gnan, T.~B. Schr{\o}der, and J.~C. Dyre.
\newblock Pressure-energy correlations in liquids. {I}. {R}esults from computer
simulations.
\newblock \emph{The Journal of Chemical Physics}, 129\penalty0 (18):\penalty0
184507, Nov 2008{\natexlab{b}}.
\newblock \doi{10.1063/1.2982247}.
\newblock URL \url{https://aip.scitation.org/doi/full/10.1063/1.2982247}.

\bibitem[Schr{\o}der et~al.(2009)Schr{\o}der, Bailey, Pedersen, Gnan, and
Dyre]{Schrder_Bailey_Pedersen_Gnan_Dyre_2009}
T.~B. Schr{\o}der, N.~P. Bailey, U.~R. Pedersen, N.~Gnan, and J.~C. Dyre.
\newblock Pressure-energy correlations in liquids. {III}. {S}tatistical
mechanics and thermodynamics of liquids with hidden scale invariance.
\newblock \emph{The Journal of Chemical Physics}, 131\penalty0 (23):\penalty0
234503, Dec 2009.
\newblock \doi{10.1063/1.3265955}.
\newblock URL \url{https://aip.scitation.org/doi/full/10.1063/1.3265955}.

\bibitem[Gnan et~al.(2009)Gnan, Schr{\o}der, Pedersen, Bailey, and
	Dyre]{Gnan_Schrder_Pedersen_Bailey_Dyre_2009}
	N.~Gnan, T.~B. Schr{\o}der, U.~R. Pedersen, N.~P. Bailey, and J.~C. Dyre.
\newblock Pressure-energy correlations in liquids. {IV}. “{I}somorphs” in
	liquid phase diagrams.
\newblock \emph{The Journal of Chemical Physics}, 131\penalty0 (23):\penalty0
	234504, Dec 2009.
\newblock \doi{10.1063/1.3265957}.
\newblock URL \url{https://aip.scitation.org/doi/full/10.1063/1.3265957}.

\bibitem[Schr{\o}der et~al.(2011)Schr{\o}der, Gnan, Pedersen, Bailey, and
	Dyre]{Schrder_Gnan_Pedersen_Bailey_Dyre_2011}
	T.~B. Schr{\o}der, N.~Gnan, U.~R. Pedersen, N.~P. Bailey, and J.~C. Dyre.
\newblock Pressure-energy correlations in liquids. {V}. {I}somorphs in
	generalized lennard-jones systems.
\newblock \emph{The Journal of Chemical Physics}, 134\penalty0 (16):\penalty0
	164505, Apr 2011.
\newblock \doi{10.1063/1.3582900}.
\newblock URL \url{https://aip.scitation.org/doi/full/10.1063/1.3582900}.

\bibitem[Schr{\o}der and Dyre(2014)]{Schrder_Dyre_2014}
	T.~B. Schr{\o}der and J.~C. Dyre.
\newblock Simplicity of condensed matter at its core: {G}eneric definition of a
	{R}oskilde-simple system.
\newblock \emph{The Journal of Chemical Physics}, 141\penalty0 (20):\penalty0
	204502, Nov 2014.
\newblock \doi{10.1063/1.4901215}.
\newblock URL \url{https://aip.scitation.org/doi/10.1063/1.4901215}.

\bibitem[Pedersen et~al.(2018)Pedersen, Schr{\o}der, and
	Dyre]{Pedersen_Schrder_Dyre_2018}
	U.~R. Pedersen, T.~B. Schr{\o}der, and J.~C. Dyre.
\newblock Phase diagram of {K}ob-{A}ndersen-type binary lennard-jones mixtures.
\newblock \emph{Physical Review Letters}, 120\penalty0 (16):\penalty0 165501,
	Apr 2018.
\newblock \doi{10.1103/PhysRevLett.120.165501}.
\newblock URL \url{https://link.aps.org/doi/10.1103/PhysRevLett.120.165501}.

\bibitem[Friisberg et~al.(2017)Friisberg, Costigliola, and
  Dyre]{Friisberg_Costigliola_Dyre_2017}
I.~M. Friisberg, L.~Costigliola, and J.~C. Dyre.
\newblock Density-scaling exponents and virial potential-energy correlation
  coefficients for the (2n, n) {L}ennard-{J}ones system.
\newblock \emph{Journal of Chemical Sciences}, 129\penalty0 (7):\penalty0
  919–928, Jul 2017.
\newblock \doi{10.1007/s12039-017-1307-1}.
\newblock URL \url{https://doi.org/10.1007/s12039-017-1307-1}.

\bibitem[Veldhorst et~al.(2015)Veldhorst, Schr{\o}der, and
  Dyre]{Veldhorst_Schrder_Dyre_2015}
A.~A. Veldhorst, T.~B. Schr{\o}der, and J.~C. Dyre.
\newblock Invariants in the {Y}ukawa system’s thermodynamic phase diagram.
\newblock \emph{Physics of Plasmas}, 22\penalty0 (7):\penalty0 073705, Jul
  2015.
\newblock \doi{10.1063/1.4926822}.
\newblock URL \url{https://aip.scitation.org/doi/10.1063/1.4926822}.

\bibitem[Bacher et~al.(2018)Bacher, Schr{\o}der, and
  Dyre]{Bacher_Schrder_Dyre_2018a}
A.~K. Bacher, T.~B. Schr{\o}der, and J.~C. Dyre.
\newblock The {EXP} pair-potential system. {II}. {F}luid phase isomorphs.
\newblock \emph{The Journal of Chemical Physics}, 149\penalty0 (11):\penalty0
  114502, Sep 2018.
\newblock \doi{10.1063/1.5043548}.
\newblock URL \url{https://aip.scitation.org/doi/10.1063/1.5043548}.

\bibitem[Veldhorst et~al.(2014)Veldhorst, Dyre, and
  Schrøder]{veldhorst_scaling_2014}
A.~A. Veldhorst, J.~C. Dyre, and T.~B. Schrøder.
\newblock Scaling of the dynamics of flexible lennard-jones chains.
\newblock \emph{The Journal of Chemical Physics}, 141\penalty0 (5):\penalty0
  054904, Aug 2014.
\newblock ISSN 0021-9606, 1089-7690.
\newblock \doi{10.1063/1.4888564}.
\newblock URL \url{http://aip.scitation.org/doi/10.1063/1.4888564}.

\bibitem[Hansen and McDonald(2013)]{hansen_mcdonald_2013}
J.~P. Hansen and I.~R. McDonald.
\newblock \emph{Theory of simple liquids: with applications of soft matter}.
\newblock Elsevier/AP, Amstersdam, fourth edition edition, 2013.
\newblock ISBN 9780123870322.

\bibitem[Roed et~al.(2013)Roed, Gundermann, Dyre, and
  Niss]{roed_communication_2013}
L.~A. Roed, D.~Gundermann, J.~C. Dyre, and K.~Niss.
\newblock Communication: Two measures of isochronal superposition.
\newblock \emph{The Journal of Chemical Physics}, 139\penalty0 (10):\penalty0
  101101, Sep 2013.
\newblock ISSN 0021-9606.
\newblock \doi{10.1063/1.4821163}.
\newblock URL \url{https://aip.scitation.org/doi/10.1063/1.4821163}.

\bibitem[Gundermann et~al.(2011)Gundermann, Pedersen, Hecksher, Bailey,
  Jakobsen, Christensen, Olsen, Schr{\o}der, Fragiadakis, and
  Casalini]{Gundermann_Pedersen_Hecksher_Bailey_Jakobsen_Christensen_Olsen_Schrder_Fragiadakis_Casalini}
D.~Gundermann, U.~R. Pedersen, T.~Hecksher, N.~P. Bailey, B.~Jakobsen,
  T.~Christensen, N.~B. Olsen, T.~B. Schr{\o}der, D.~Fragiadakis, and
  R.~Casalini.
\newblock Predicting the density-scaling exponent of a glass-forming liquid
  from prigogine–defay ratio measurements.
\newblock \emph{Nature Physics}, 7\penalty0 (10):\penalty0 816–821, 2011.
\newblock \doi{10.1038/nphys2031}.
\newblock URL \url{http://www.nature.com/articles/nphys2031}.

\bibitem[Hansen et~al.(2018)Hansen, Sanz, Adrjanowicz, Frick, and
  Niss]{Hansen_Sanz_Adrjanowicz_Frick_Niss_2018}
H.~W. Hansen, A.~Sanz, K.~Adrjanowicz, B.~Frick, and K.~Niss.
\newblock Evidence of a one-dimensional thermodynamic phase diagram for simple
  glass-formers.
\newblock \emph{Nature Communications}, 9\penalty0 (1):\penalty0 1–7, Feb
  2018.
\newblock \doi{10.1038/s41467-017-02324-3}.
\newblock URL \url{https://www.nature.com/articles/s41467-017-02324-3}.

\bibitem[Xiao et~al.(2015)Xiao, Tofteskov, Christensen, Dyre, and
  Niss]{XIAO2015190}
W. Xiao, J. Tofteskov, T.~V. Christensen, J.~C. Dyre, and K.
  Niss.
\newblock Isomorph theory prediction for the dielectric loss variation along an
  isochrone.
\newblock \emph{Journal of Non-Crystalline Solids}, 407:\penalty0 190--195,
  2015.
\newblock ISSN 0022-3093.
\newblock \doi{https://doi.org/10.1016/j.jnoncrysol.2014.08.041}.
\newblock URL
  \url{https://www.sciencedirect.com/science/article/pii/S0022309314004268}.
\newblock 7th IDMRCS: Relaxation in Complex Systems.

\bibitem[Dyre(2016)]{dyre_simple_2016}
J.~C Dyre.
\newblock Simple liquids’ quasiuniversality and the hard-sphere paradigm.
\newblock \emph{Journal of Physics: Condensed Matter}, 28\penalty0
  (32):\penalty0 323001, Aug 2016.
\newblock ISSN 0953-8984, 1361-648X.
\newblock \doi{10.1088/0953-8984/28/32/323001}.
\newblock URL
  \url{https://iopscience.iop.org/article/10.1088/0953-8984/28/32/323001}.

\bibitem[Dyre(2018)]{dyre_perspective_2018}
J.~C. Dyre.
\newblock Perspective: Excess-entropy scaling.
\newblock \emph{The Journal of Chemical Physics}, 149\penalty0 (21):\penalty0
  210901, Dec 2018.
\newblock ISSN 0021-9606.
\newblock \doi{10.1063/1.5055064}.
\newblock URL \url{https://aip.scitation.org/doi/10.1063/1.5055064}.

\bibitem[Costigliola et~al.(2016)Costigliola, Schr{\o}der, and
  Dyre]{Costigliola_Schrder_Dyre_2016}
L.~Costigliola, T.~B. Schr{\o}der, and J.~C. Dyre.
\newblock Freezing and melting line invariants of the {L}ennard-{J}ones system.
\newblock \emph{Physical Chemistry Chemical Physics}, 18\penalty0
  (21):\penalty0 14678–14690, 2016.
\newblock \doi{10.1039/C5CP06363A}.
\newblock URL \url{http://xlink.rsc.org/?DOI=C5CP06363A}.

\bibitem[Heyes et~al.(2019)Heyes, Dini, Costigliola, and
  Dyre]{heyes_transport_2019}
D.~M. Heyes, D.~Dini, L.~Costigliola, and J.~C. Dyre.
\newblock Transport coefficients of the lennard-jones fluid close to the
  freezing line.
\newblock \emph{The Journal of Chemical Physics}, 151\penalty0 (20):\penalty0
  204502, Nov 2019.
\newblock ISSN 0021-9606.
\newblock \doi{10.1063/1.5128707}.
\newblock URL \url{https://aip.scitation.org/doi/10.1063/1.5128707}.

\bibitem[Bailey et~al.(2017)Bailey, Ingebrigtsen, Hansen, Veldhorst, Bøhling,
  Lemarchand, Olsen, Bacher, Costigliola, Pedersen, Larsen, Dyre, and
  Schrøder]{bailey_rumd_2017}
N.~P. Bailey, T.~S. Ingebrigtsen, J.~S. Hansen, A.~A. Veldhorst, L.~Bøhling,
  C.~A. Lemarchand, A.~E. Olsen, A.~K. Bacher, L.~Costigliola, U.~R. Pedersen,
  H.~Larsen, J.~C. Dyre, and T.~B. Schrøder.
\newblock {RUMD}: A general purpose molecular dynamics package optimized to
  utilize {GPU} hardware down to a few thousand particles.
\newblock \emph{{SciPost} Physics}, 3\penalty0 (6):\penalty0 038, Dec 2017.
\newblock ISSN 2542-4653.
\newblock \doi{10.21468/SciPostPhys.3.6.038}.
\newblock URL \url{http://arxiv.org/abs/1506.05094}.

\bibitem[Allen and Tildesley(2017)]{allen_computer_2017}
M.~P. Allen and D.~J. Tildesley.
\newblock \emph{Computer simulation of liquids}.
\newblock Oxford University Press, second edition edition, 2017.
\newblock ISBN 9780198803195 9780198803201.

\bibitem[Hansen(2021{\natexlab{a}})]{hansenNEWbook}
J.~S. Hansen.
\newblock \emph{Aspects of Nanofluidics, to appear}.
\newblock Cambridge University Press, 2021{\natexlab{a}}.

\bibitem[Hansen et~al.(2015)Hansen, Dyre, Daivis, Todd, and
  Bruus]{Hansen_Dyre_Daivis_Todd_Bruus_2015}
J.~S. Hansen, J.~C. Dyre, P.~Daivis, B.~D. Todd, and H.~Bruus.
\newblock Continuum nanofluidics.
\newblock \emph{Langmuir}, 31\penalty0 (49):\penalty0 13275–13289, Dec 2015.
\newblock \doi{10.1021/acs.langmuir.5b02237}.
\newblock URL \url{https://pubs.acs.org/doi/10.1021/acs.langmuir.5b02237}.

\bibitem[Todd and Daivis(2017)]{Todd_Daivis_book}
B.~D. Todd and P.~J. Daivis.
\newblock \emph{Nonequilibrium Molecular Dynamics: Theory, Algorithms and
  Applications}.
\newblock Cambridge University Press, 2017.

\bibitem[Irving and Kirkwood(1950)]{irving_kirkwood_1950}
J.~H. Irving and J.~G. Kirkwood.
\newblock The {Statistical} {Mechanical} {Theory} of {Transport} {Processes}.
  {IV}. {The} {Equations} of {Hydrodynamics}.
\newblock \emph{The Journal of Chemical Physics}, 18\penalty0 (6):\penalty0
  817--829, Jun 1950.
\newblock ISSN 0021-9606.
\newblock \doi{10.1063/1.1747782}.
\newblock URL \url{https://aip.scitation.org/doi/10.1063/1.1747782}.

\bibitem[Pedersen et~al.(2008)Pedersen, Bailey, Schrøder, and
  Dyre]{pedersen_strong_2008}
U.~R. Pedersen, N.~P. Bailey, T.~B. Schrøder, and J.~C. Dyre.
\newblock Strong {Pressure}-{Energy} {Correlations} in van der {Waals}
  {Liquids}.
\newblock \emph{Physical Review Letters}, 100\penalty0 (1):\penalty0 015701,
  Jan 2008.
\newblock ISSN 0031-9007, 1079-7114.
\newblock \doi{10.1103/PhysRevLett.100.015701}.
\newblock URL \url{https://link.aps.org/doi/10.1103/PhysRevLett.100.015701}.

\bibitem[Pedersen et~al.(2016)Pedersen, Costigliola, Bailey, Schrøder, and
  Dyre]{pedersen_thermodynamics_2016}
U.~R. Pedersen, L. Costigliola, N.~P. Bailey, T.~B. Schrøder,
  and J.~C. Dyre.
\newblock Thermodynamics of freezing and melting.
\newblock \emph{Nature Communications}, 7\penalty0 (1):\penalty0 12386,
  Nov 2016.
\newblock ISSN 2041-1723.
\newblock \doi{10.1038/ncomms12386}.
\newblock URL \url{http://www.nature.com/articles/ncomms12386}.

\bibitem[Ingebrigtsen et~al.(2012)Ingebrigtsen, Bøhling, Schrøder, and
  Dyre]{ingebrigtsen_communication_2012}
T.~S. Ingebrigtsen, L. Bøhling, T.~B. Schrøder, and J.~C. Dyre.
\newblock Communication: {Thermodynamics} of condensed matter with strong
  pressure-energy correlations.
\newblock \emph{The Journal of Chemical Physics}, 136\penalty0 (6):\penalty0
  061102, Feb 2012.
\newblock ISSN 0021-9606.
\newblock \doi{10.1063/1.3685804}.
\newblock URL \url{https://aip.scitation.org/doi/10.1063/1.3685804}.

\bibitem[Bøhling et~al.(2012)Bøhling, Ingebrigtsen, Grzybowski, Paluch, Dyre,
  and Schrøder]{bohling_scaling_2012}
L.~Bøhling, T.~S. Ingebrigtsen, A.~Grzybowski, M.~Paluch, J.~C. Dyre, and T.~B.
  Schrøder.
\newblock Scaling of viscous dynamics in simple liquids: theory, simulation and
  experiment.
\newblock \emph{New Journal of Physics}, 14\penalty0 (11):\penalty0 113035,
  Nov 2012.
\newblock ISSN 1367-2630.
\newblock \doi{10.1088/1367-2630/14/11/113035}.
\newblock URL
  \url{https://iopscience.iop.org/article/10.1088/1367-2630/14/11/113035}.

\bibitem[Furukawa and Tanaka(2009)]{furukawa_nonlocal_2009}
A.~Furukawa and H.~Tanaka.
\newblock Nonlocal nature of the viscous transport in supercooled liquids:
  Complex fluid approach to supercooled liquids.
\newblock \emph{Physical Review Letters}, 103\penalty0 (13):\penalty0 135703,
  Sep 2009.
\newblock \doi{10.1103/PhysRevLett.103.135703}.
\newblock URL \url{https://link.aps.org/doi/10.1103/PhysRevLett.103.135703}.

\bibitem[Martin and Siepmann(1998)]{martin_transferable_1998}
M.~G. Martin and J.~I. Siepmann.
\newblock Transferable potentials for phase equilibria. 1. {U}nited-atom
  description of n-alkanes.
\newblock \emph{The Journal of Physical Chemistry B}, 102\penalty0
  (14):\penalty0 2569--2577, Apr 1998.
\newblock ISSN 1520-6106.
\newblock \doi{10.1021/jp972543+}.
\newblock URL \url{https://doi.org/10.1021/jp972543+}.

\bibitem[Hansen et~al.(2007)Hansen, Daivis, Travis, and
  Todd]{hansen_parameterization_2007}
J.~S. Hansen, P.~J. Daivis, K.~P. Travis, and B.~D. Todd.
\newblock Parameterization of the nonlocal viscosity kernel for an atomic
  fluid.
\newblock \emph{Physical Review E}, 76\penalty0 (4):\penalty0 041121, Oct 2007.
\newblock \doi{10.1103/PhysRevE.76.041121}.
\newblock URL \url{https://link.aps.org/doi/10.1103/PhysRevE.76.041121}.

\bibitem[Costigliola et~al.(2018)Costigliola, Pedersen, Heyes, Schrøder, and
  Dyre]{costigliola_communication_2018}
L.~Costigliola, U.~R. Pedersen, D.~M. Heyes, T.~B. Schrøder, and J.~C. Dyre.
\newblock Communication: {S}imple liquids’ high-density viscosity.
\newblock \emph{The Journal of Chemical Physics}, 148\penalty0 (8):\penalty0
  081101, Feb 2018.
\newblock ISSN 0021-9606.
\newblock \doi{10.1063/1.5022058}.
\newblock URL \url{https://aip.scitation.org/doi/10.1063/1.5022058}.

\bibitem[Puscasu et~al.(2010)Puscasu, Todd, Daivis, and
  Hansen]{puscasu_nonlocal_2010}
R.~M. Puscasu, B.~D. Todd, P.~J. Daivis, and J.~S. Hansen.
\newblock Nonlocal viscosity of polymer melts approaching their glassy state.
\newblock \emph{The Journal of Chemical Physics}, 133\penalty0 (14):\penalty0
  144907, Oct 2010.
\newblock ISSN 0021-9606.
\newblock \doi{10.1063/1.3499745}.
\newblock URL \url{https://aip.scitation.org/doi/10.1063/1.3499745}.

\bibitem[Phan-Thien and Mai-Duy(2017)]{phan-thien_understanding_2017}
N.~Phan-Thien and N.~Mai-Duy.
\newblock \emph{Understanding Viscoelasticity: {A}n Introduction to Rheology}.
\newblock Graduate Texts in Physics. Springer International Publishing, 2017.
\newblock ISBN 9783319619996 9783319620008.
\newblock \doi{10.1007/978-3-319-62000-8}.
\newblock URL \url{http://link.springer.com/10.1007/978-3-319-62000-8}.

\bibitem[Trachenko and Brazhkin(2016)]{trachenko_collective_2016}
K.~Trachenko and V.~V. Brazhkin.
\newblock Collective modes and thermodynamics of the liquid state.
\newblock \emph{Reports on Progress in Physics}, 79\penalty0 (1):\penalty0
  016502, Jan 2016.
\newblock ISSN 0034-4885, 1361-6633.
\newblock \doi{10.1088/0034-4885/79/1/016502}.
\newblock URL
  \url{https://iopscience.iop.org/article/10.1088/0034-4885/79/1/016502}.

\bibitem[Dyre(2013)]{dyre_isomorphs_2013}
J.~C. Dyre.
\newblock Isomorphs, hidden scale invariance, and quasiuniversality.
\newblock \emph{Physical Review E}, 88\penalty0 (4):\penalty0 042139, Oct 2013.
\newblock \doi{10.1103/PhysRevE.88.042139}.
\newblock URL \url{https://link.aps.org/doi/10.1103/PhysRevE.88.042139}.

\bibitem[Hansen(2021{\natexlab{b}})]{JSHmolsim}
J.~S. Hansen.
\newblock Where is the hydrodynamic limit?
\newblock \emph{Molecular Simulation}, 0\penalty0 (0):\penalty0 1--11,
  2021{\natexlab{b}}.
\newblock \doi{10.1080/08927022.2021.1975038}.
\newblock URL \url{https://doi.org/10.1080/08927022.2021.1975038}.

\bibitem[Bell et~al.(2020)Bell, Dyre, and
  Ingebrigtsen]{bell_excess-entropy_2020}
I.~H. Bell, J.~C. Dyre, and T.~S. Ingebrigtsen.
\newblock Excess-entropy scaling in supercooled binary mixtures.
\newblock \emph{Nature Communications}, 11\penalty0 (1):\penalty0 4300, Aug
  2020.
\newblock ISSN 2041-1723.
\newblock \doi{10.1038/s41467-020-17948-1}.
\newblock URL \url{https://www.nature.com/articles/s41467-020-17948-1}.

\end{thebibliography}
\end{document}